\documentclass[useAMS,usenatbib]{mn2e}

\usepackage{amsmath,array,amsfonts}
\usepackage{amssymb}
\usepackage{graphicx}
\usepackage{placeins}
\usepackage{float}
\usepackage{algorithm}
\usepackage[noend]{algpseudocode}
\usepackage{natbib}
\usepackage{pstricks}
\usepackage{verbatim}
\usepackage{subfigure}
\usepackage{booktabs}
\usepackage{siunitx,booktabs,multirow, dcolumn}

\newcommand{\ltsima} {$\; \buildrel < \over \sim \;$}
\newcommand{\gtsima} {$\; \buildrel > \over \sim \;$}
\newcommand{\lta} {\lower.5ex\hbox{\ltsima}}
\newcommand{\gta} {\lower.5ex\hbox{\gtsima}}

\newcolumntype{d}[1]{D{.}{.}{#1} }

\def\f{\frac}
\def\nn{\nonumber}

\def\exp{\mathrm{exp}}
\def\cov{\mathbf{C}}
\def\noise{\mathbf{N}}
\def\isotropic{\mathbf{T}}
\def\aniso{\bar{\mathbf{N}}}
\def\field{\mathbf{s}}
\def\data{\mathbf{d}}
\def\messenger{\mathbf{t}}
\def\bs{\boldsymbol}

\begin{document}

\title{Hierarchical Cosmic Shear Power Spectrum Inference}

\author[J. Alsing, A. Heavens, A. Jaffe, A. Kiessling, B. Wandelt, T. Hoffmann]
{\parbox{\textwidth}{Justin Alsing$^1$\thanks{e-mail:  j.alsing12@imperial.ac.uk}, 
Alan Heavens$^1$, 
Andrew H. Jaffe$^1$,
Alina Kiessling$^2$,
Benjamin Wandelt$^{3,4}$ and
Till Hoffmann$^5$}\vspace{0.4cm}\\
\parbox{\textwidth}{$^1$ Imperial Centre for Inference and Cosmology, Department of Physics, Imperial College London, Blackett Laboratory, Prince Consort Road, London SW7 2AZ, UK\\
$^2$ Jet Propulsion Laboratory, California Institute of Technology, 4800 Oak Grove Drive, Pasadena, CA 91109, USA\\
$^3$ Institut d'Astrophysique de Paris, UMR CNRS 7095, Universit\'{e} Pierre et Marie Curie, 98bis boulevard Arago, 75014 Paris, France\\
$^4$ Department of Physics, University of Illinois at Urbana-Champaign, Urbana, Illinois 61801, USA\\
$^5$ Department of Mathematics, South Kensington Campus, Imperial College London, London SW7 2AZ, UK}}
\date{Accepted ;  Received ; in original form }

\maketitle

\begin{abstract}
We develop a Bayesian hierarchical modelling approach for cosmic shear power spectrum inference, jointly sampling from the posterior distribution of the cosmic shear field and its (tomographic) power spectra. Inference of the shear power spectrum is a powerful intermediate product for a cosmic shear analysis, since it requires very few model assumptions and can be used to perform inference on a wide range of cosmological models \emph{a posteriori} without loss of information. We show that joint posterior for the shear map and power spectrum can be sampled effectively by Gibbs sampling, iteratively drawing samples from the map and power spectrum, each conditional on the other. This approach neatly circumvents difficulties associated with complicated survey geometry and masks that plague frequentist power spectrum estimators, since the power spectrum inference provides prior information about the field in masked regions at every sampling step. We demonstrate this approach for inference of tomographic shear $E$-mode, $B$-mode and $EB$-cross power spectra from a simulated galaxy shear catalogue with a number of important features; galaxies distributed on the sky and in redshift with photometric redshift uncertainties, realistic random ellipticity noise for every galaxy and a complicated survey mask. The obtained posterior distributions for the tomographic power spectrum coefficients recover the underlying simulated power spectra for both $E$- and $B$-modes.

\end{abstract}
\section{Introduction}
\label{intro}
As light from distant galaxies propagates through the Universe it is continuously deflected by the gravitational potential of the large-scale matter distribution, resulting in a coherent distortion of observed galaxy images on the sky. This weak gravitational lensing provides a powerful probe of the growth rate of potential perturbations and the geometry of the Universe through the distance-redshift relation. Weak lensing has unique appeal in its sensitivity to the full matter distribution in the Universe, allowing us to probe the 3D matter power spectrum over a range of scales and redshifts. Since weak lensing is a function of both the geometry of the universe and the growth of structure, it is a particularly sensitive probe of dark energy and gravity on large scales (see e.g., \citealp{Weinberg2013} and references therein). Cosmic shear analyses in particular aim to extract cosmological information from weak lensing by measuring correlations of galaxy ellipticities, modified by the lensing field, across the sky (see \citealp{Munshi2008} for a comprehensive review). The natural starting point for such an analysis is to look at the two-point statistics of the shear field, although a substantial amount of information may also be available in the higher-order shear statistics \citep{Bernardeau1997, VanWaerbeke1999, Schneider2003, Takada2003, Vafaei2010, Kayo2013}. This paper is concerned with extracting cosmological inferences from the two-point statistics of the cosmic shear field.

When analysing the two-point statistics of a random field, we are free to choose the most convenient basis to work in. For example, working in the pixel basis the two-point function of the shear field is the real-space correlation function, whilst if we choose to work in harmonic space the two-point function is the angular power spectrum. The power spectrum has a clear advantage over the correlation function; due to the statistical isotropy of the shear field, its spherical harmonic coefficients are uncorrelated and hence the covariance matrix of the field in this basis is sparse. The covariance of the real-space shear field on the other hand is not sparse, since the correlation function is non-zero over all scales. This is a major advantage of using the power spectrum over the correlation function in cosmic shear analyses and will become increasingly important for larger-area weak lensing surveys.

Inference of the shear power spectrum is an incredibly powerful intermediate product for a cosmic shear analysis, since it can be used to perform inference on a wide range of cosmological models \emph{a posteriori} without loss of information. Power spectrum inference is almost model independent, assuming only that the field under inspection is statistically isotropic. In this study we further assume that the field is well-described by Gaussian statistics, but also note that the Gaussian distribution constitutes the maximum entropy prior once the mean and covariance are specified, so from a Bayesian perspective assuming a Gaussian distribution for the field may be a well-justified (although sub-optimal) approximation even on small scales where the shear field is non-linear. Since a wide range of cosmological models provide a deterministic relationship between the cosmological model parameters and the power spectrum, cosmological parameter inference can be performed \emph{a posteriori} from the posterior distribution of the power spectrum given the data, without loss of information. This is clearly preferable to performing inference for each model independently from the full data-set and allows for efficient analysis of future cosmological models without having to re-analyse the full data set from scratch.

In this paper we develop a Bayesian method for inferring the cosmic shear power spectrum by jointly sampling from the posterior distribution of the shear map and power spectrum given the data, in a hierarchical fashion. A similar approach has been developed for analysing cosmic microwave background (CMB) temperature maps \citep{Wandelt2004, Eriksen2004, Odwyer2004, Chu2005, Komatsu2011, Planck2013XV}, CMB polarization \citep{Larson2007, Eriksen2007, Komatsu2011, Karakci2013} and large-scale structure \citep{Jasche2010, Jasche2012, Jasche2013}. Cosmic shear bears some similarities and some important differences to the CMB and large-scale structure inference problems. The shear field is a 3D random field with spin-weight-2 on the angular sky. Shear is generally assumed to be a statistically isotropic field, but since lensing is an integrated effect it is not statistically homogeneous, in the sense that its statistical properties evolve strongly with redshift. On large scales, the shear field is Gaussian to a good approximation, whilst on small scales it becomes significantly non-Gaussian. As such, weak lensing combines together many of the key features seen in the context of CMB temperature, polarization and LSS power spectrum inference problems, whilst the inhomogeneity of the shear field adds an additional new feature; in this sense weak lensing power spectrum inference is a particularly rich statistical problem.

One of the main challenges in estimating the power spectrum from weak lensing survey data is accounting for complicated survey geometry due to both incomplete sky coverage and masked regions within the survey area. The problem stems from the fact that Fourier and spherical harmonic basis functions are not orthogonal on the cut sky, which can result in masks moving power from the angular scale of the masks to other parts of the power spectrum and leakage between $E$- and $B$-mode power. These problems are a major inconvenience for approximate power spectrum estimation methods such as the pseudo-$C_\ell$ (see e.g., \citealp{Chon2004} and \citealp{Brown2005}, and also \citealp{Smith2006} for an estimator approach that avoids some of these issues). Fortunately the approach presented here bypasses these difficulties completely. By estimating the map and power spectrum simultaneously in a block-MCMC or Gibbs sampling framework, we iteratively sample from the map conditional on the power spectrum and the power spectrum conditional on the map. When sampling from the conditional distribution of the map with a fixed power spectrum, even though the data provides no information about the masked regions the power spectrum still provides (probabilistic) information about the field in those regions. Masked regions are treated as pixels with infinite noise, but the inferred power spectrum (combined with inference of pixels surrounding the masked region) nonetheless informs us about the field in the masked regions, circumventing the need to treat masked regions as being cut from the analysis and simplifying the survey geometry.

Drawing inferences about cosmology from weak lensing survey data is a challenging task, involving a number of complex modelling elements. Measuring galaxy shapes and redshifts from pixelized images and photometric data requires detailed models for the telescope point-spread function (PSF), seeing effects, pixel noise and other instrumental effects, as well as models for the intrinsic distributions of galaxy properties which determine their physical and photometric appearance. Cosmological parameter inference from observed galaxy shapes and redshifts requires a model relating the cosmology (statistically) to the cosmic shear field and in turn its impact on observed galaxies. Formulating the weak lensing inference task as a global hierarchical model is an attractive approach for a number of reasons. Hierarchical models account for the full statistical interdependency structure of all model components and allow information to flow freely from raw pixel and photometric data through to cosmological inferences; this is optimal in the sense that no information is lost, and principled in the sense that parameter uncertainties are propagated correctly and completely throughout the analysis (see \citealp{Schneider2015} for a discussion). However, whilst it is the optimal approach in principle, in practice performing a global analysis on a data set as large and complex as a weak lensing survey is a formidable challenge and a global analysis may not be computationally feasible. The alternative approach is to break the global problem up into a number of sub-problems which are analysed in a series of steps, where the output of each step is used as input for the next. Whilst this has the advantage that sub-problems are computationally easier to solve, it is sub-optimal in the sense that full parameter interdependencies are not accounted for and it is challenging to propagate uncertainties consistently throughout the pipeline, introducing biases which must be carefully corrected for. The hierarchical modelling approach to map-power spectrum inference developed here will form a central part of any larger global hierarchical model for weak lensing, or alternatively Bayesian map-power spectrum inference can be performed in isolation from e.g., a catalogue of measured galaxy shapes and redshifts, once instrumental effects, ellipticity distributions etc have been modelled and accounted for \emph{a priori}. 

The structure of this paper is as follows: in \S \ref{sec:hierarchical_wl} we discuss the global hierarchical modelling approach to cosmic shear inference and isolate the map-power spectrum inference problem in \S \ref{sec:isolating}. In \S \ref{sec:tomography} we specialize to tomographic cosmic shear and in \S \ref{sec:map_power_inference} we develop the Gibbs sampling approach for joint shear map-power spectrum inference. In \S \ref{sec:sims} we describe the shear simulations and we demonstrate the method by recovering tomographic $E$- and $B$-mode shear power spectra from a simulated shear catalogue in \S \ref{sec:results}. We conclude in \S \ref{sec:conclusions}.

\section{Weak Lensing as a hierarchical modelling problem}
\label{sec:hierarchical_wl}
The statistical task in weak lensing can be summarized as follows: given noisy, PSF convolved, pixelized images and photometric data for a large number of galaxies, how do we make inferences about cosmological models and parameters? The full inference problem involves modelling a number of processes, which can be broadly split into three groups: (1) How is the shear field (statistically) related to the cosmological model/parameters? (2) What are the distributions of galaxy properties that characterise their appearance, both those which change under lensing (e.g., shape, size, apparent magnitude etc) and those which do not (e.g., S\'{e}rsic index, colour etc), and what is the distribution of galaxy redshifts? (3) What are the telescope PSF, noise properties and other instrumental effects that generate noisy, pixelized, PSF convolved images and photometric data from galaxies with given physical characteristics and redshifts? In principle we can write down a hierarchical model combining all of these processes and solve the global inference problem. In this section we outline the global hierarchical modelling approach for weak lensing and highlight the key advantages and challenges. This work complements that of \citet{Schneider2015}, who investigated in some detail the sections of the hierarchy concerned with pixel data and galaxy properties. Here we focus on a different part of the hierarchy, starting with point estimates of shear, and ending with the shear power spectra.  A global statistical model would include both. 
\begin{figure}
\begin{center}
\includegraphics[width = 9cm]{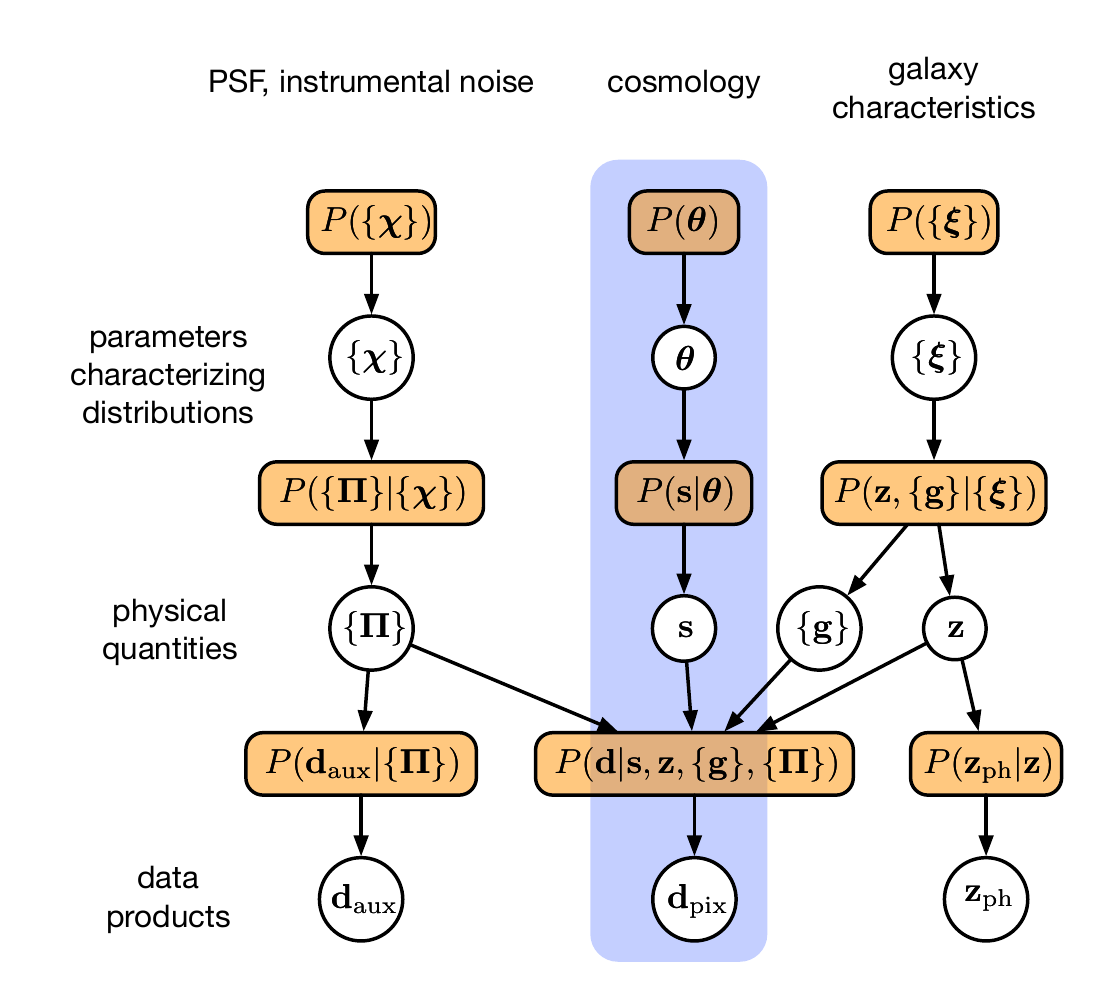}
\end{center}
\caption{A generative forward model for weak lensing pixel and photometric data: cosmological parameters $\bs{\theta}$, a set of parameters $\{\bs{\xi}\}$ characterizing the distributions of physical galaxy characteristics and redshifts and parameters $\{\bs{\chi}\}$ characterizing the distribution of PSF and instrumental noise properties are realised from their respective prior distributions. The cosmology then generates a realization of the shear field $\field$, a set of physical galaxy properties $\{\mathbf{g}\}$ and redshifts $\mathbf{z}$ are generated for each galaxy from their specified distributions, and a set of effective PSFs and instrumental noise properties (for each band, epoch etc) $\{\bs{\Pi}\}$ are realised given their intrinsic distributions characterized by $\{\bs{\chi}\}$. Pixelized images for each galaxy $\data_\mathrm{pix}$ are then realised given the galaxy characteristics, redshifts, instrumental PSF and noise properties, photometric data $\mathbf{z}_\mathrm{ph}$ (on which photometric redshift inference is based) are generated given the true redshifts, and some auxiliary data $\data_\mathrm{aux}$ providing additional information about the instrumental properties is realised given the PSF and noise properties. Note that this is by no means the most general model and might straightforwardly be extended to include additional parameter and data interdependencies, additional model parameters, hyperpriors and additional data products.}
\label{fig:uber_model}
\end{figure}

In order to write down a global hierarchical model for weak lensing let us first define a set of parameters that the model could include. A basic but comprehensive set of model parameters could be defined as follows: cosmological parameters $\bs{\theta}$, the shear field $\field$, a set of physical galaxy properties (size, shape, magnitude etc) $\{\mathbf{g}\}$ and redshifts $\mathbf{z}$ for each galaxy, a set of parameters $\{\bs{\xi}\}$ characterizing the distributions of intrinsic galaxy properties and redshifts, a set of parameters $\{\bs{\chi}\}$ characterizing the distribution of PSF parameters and pixel noise and a set of parameters $\{\bs{\Pi}\}$ characterizing the effective PSF and noise for different photometric bands, epochs, positions on the sky etc. The data are pixelized images for each galaxy $\data_\mathrm{pix}$, photometric data for each source (on which the photometric redshift inference is based) $\mathbf{z}_\mathrm{ph}$ and some auxiliary data providing additional information specifically about the PSF and pixel noise properties $\data_\mathrm{aux}$. Note that the discussion in this section is quite general and the field $\field$ refers to the full 3D shear field; in \S \ref{sec:tomography} we specialise to tomographic shear where $\field$ will thereafter refer to the set of 2D tomographic shear fields.

A particularly elegant and useful way of visualizing forward hierarchical models is as a directed, acyclic bipartite graph, where model parameters and data (nodes, represented by white circles) are connected via conditional probability distributions (represented by orange boxes), for example Fig. \ref{fig:uber_model} (described in detail below). This representation clearly elicits the conditional structure of the inference problem; parameters and data which are directly connected (via a single conditional density) are dependent, whereas parameters and data which are not directly connected are conditionally independent. The posterior distribution for the full set of model parameters is straightforwardly obtained in a readily factorized form, accounting for the full conditional structure of the problem, by taking the product of all distributions appearing in the graph (upto a normalization constant). Furthermore, the distribution of a single parameter node conditional on all others is given by the product of the conditional densities on all incoming and outgoing edges for that parameter (again upto a normalization constant). Each conditional probability distribution can be thought of as a separate modelling step; the hierarchical model thus breaks up the global problem into a number of sub-models, where one only needs to be able to write down conditional distributions for the various sub-sets of model parameters with all other parameters held fixed. Whilst the hierarchical model describes the global inference problem, it is still modular in the sense that the model neatly factorizes into a set of sub-problems that can be attacked in turn.

A generative forward model for the data and model assumptions described in the previous paragraphs is given in Fig. \ref{fig:uber_model} and can be understood as follows: In the top level, $\bs{\theta}$, $\{\bs{\xi}\}$ and $\{\bs{\chi}\}$ are drawn from their respective priors. In the second level, the cosmology generates a shear field $\field$, a set of intrinsic galaxy properties $\{\mathbf{g}\}$ and redshifts is generated from $P(\mathbf{z}, \{\mathbf{g}\}|\{\boldsymbol{\xi}\})$, and the PSFs and pixel noise characteristics are generated from $P(\{\boldsymbol{\Pi}\}| \{\boldsymbol{\chi}\})$. Finally, the pixel values are generated given the galaxy positions and physical properties, the shear field, PSFs and pixel noise characteristics, the auxiliary data is generated from $\{\boldsymbol{\Pi}\}$, and the photometric redshift data is generated from the true redshifts. This model is by no means the most general and can easily be extended to include further levels of sophistication, such as additional parameter and data interdependencies, additional model parameters, parameterized hyperpriors and additional data products.

The hierarchical model in Fig. \ref{fig:uber_model} clearly shows the statistical interdependencies between the various parts of the model. Ideally we would like to solve the global inference problem, simultaneously inferring all of the model parameters given the data and marginalizing over all latent parameters that are not of direct interest. This approach correctly accounts for the complicated web of interdependencies and allows information to flow unimpeded from the data through to the cosmological parameter inference; it is optimal in the sense that no information is lost, and principled in the sense that the uncertainties in all parameters are correctly and completely propagated throughout the analysis. In contrast, the frequentist approach typically analyses each part of the model in a series of consecutive steps, where the results of each step are used as inputs for the next. This makes it challenging to both correctly account for the full statistical interdependency and to propagate uncertainties consistently, leading to biases that must be carefully (and painstakingly) corrected for. In spite of being a global analysis, the hierarchical approach is in fact naturally modular; the Gibbs or block-MCMC sampling scheme separates the various steps in the inference process, iteratively dealing with each component in turn. Whilst the typical frequentist approach estimates fixed values for parameters at each step and feeds them into the next, the hierarchical approach feeds the full probabilistic inference about each parameter throughout the analysis. 

In principle, it is possible to perform an (asymptotically) optimal frequentist analysis, writing down the global multi-parameter likelihood and performing a joint maximum-likelihood analysis for all model parameters from the global likelihood. However, estimating parameter uncertainties accurately in a high-dimensional multi-parameter maximum-likelihood analysis is expected to be considerably more computationally challenging than implementing a Markov chain for the equivalent Bayesian inference problem, given a fixed data set. The Bayesian hierarchical modelling approach to weak lensing has clear advantages over both the sub-optimal frequentist approach, analysing each part of the model in series (as described above), and the more careful global maximum-likelihood analysis.

Given the scale and complexity of the weak lensing inference problem described in Fig. \ref{fig:uber_model}, a sensible approach to developing a global analysis pipeline is to temporarily break up the global model into a number of sub-models (for example, the three columns in Fig. \ref{fig:uber_model}), allowing us to tackle the technical challenges associated with different parts of the hierarchy in isolation. These can later be re-united to build as global an analysis pipeline as possible.

\subsection{Isolating the map-power spectrum inference problem}
\label{sec:isolating}
\begin{figure}
\begin{center}
\includegraphics{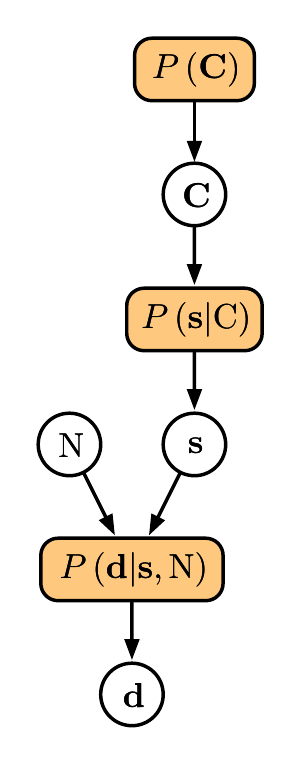}
\end{center}
\caption{Hierarchical forward model for noisy pixelized shear maps $\data$ from the tomograhic shear power spectra $\cov$: the shear power spectrum $\cov$ is drawn from some prior distribution, a realization of the tomographic shear fields $\field$ are then generated given the power spectra, and finally noisy tomographic shear maps $\data$ are realised by adding noise with covariance $\noise$.}
\label{fig:vanilla_model}
\end{figure}

In this paper we are interested in isolating the central column of the hierarchy in Fig. \ref{fig:uber_model} (highlighted in blue), i.e., extracting cosmological parameter inferences from weak lensing data via the cosmic shear field. Rather than attempt to infer cosmological parameters directly, we would rather like to infer the shear power spectrum given the data. Cosmological models provide a deterministic relationship between cosmological parameters and the shear power spectrum. This means that cosmological parameter inference can be performed \emph{a posteriori} for a wide range of models directly from the posterior distribution for the shear power spectrum given the data, without loss of information (see \S \ref{sec:br} for details). As such, inference of the shear power spectrum is a very useful intermediate product and is preferable to doing cosmological parameter inference directly from the data for the various models of interest separately. With this in mind we isolate and tackle the following sub-problem: given some (pre-processed) weak lensing data products, we want to jointly infer the shear field and its power spectrum. Furthermore, we will assume the shear field is Gaussian and fully characterized by its two-point statistics, i.e., its power spectrum (covariance matrix) $\cov$. This reduced problem is summarized by the forward model in Fig. \ref{fig:vanilla_model} and can be understood as follows: We begin by specifying a prior for the power spectrum $P(\cov)$ which generates a power spectrum $\cov$. This power spectrum then generates a shear field $\field$ via the density $P(\field|\cov)$, which we take to be a zero mean Gaussian with covariance $\cov$. We then fix the noise covariance matrix $\noise$ and add noise to the realised shear map to give a realisation of the data, a noisy estimate of the shear field, via the conditional density $P(\data|\field, \noise)$. In \S \ref{sec:tomography}, we specialize to tomography, and $\cov$ and $\field$ will be understood to denote the tomographic power spectra and tomographic fields respectively. 

Since the left and right hand sides of the hierarchy from Fig. \ref{fig:uber_model} have been removed in the reduced problem, we are forced to use as our data vector some processed version of the raw pixel data $\data_\mathrm{pix}$ for which the instrumental effects, distributions of galaxy properties and photometric redshifts have already been modelled and accounted for \emph{a priori}. We take our pre-processed data to be a catalogue of measured galaxy shapes, angular positions and photometric redshifts, where the effects of $\{\bs{\chi}\}$, $\{\bs{\Pi}\}$, $\{\bs{\xi}\}$, $\{\mathbf{g}\}$ and $\mathbf{z}$ have been accounted for in the shape and photometric redshift inference process. In the next section we specialize the problem further to a tomographic cosmic shear analysis; we will see that this requires the data to be compressed further by grouping sources into a number of redshift bins and angular pixels on the sky. The resulting data vector will be a set of noisy pixelized shear maps for multiple tomographic redshift bins.

\subsection{Tomographic cosmic shear}
\label{sec:tomography}
Ideally we would like to analyse the full 3D shear field, and formalism for performing 3D cosmic shear analysis is now well developed \citep{Heavens2003, Castro2005, Heavens2006, Kitching2007, Kitching2008a, Kitching2011, Kitching20143D}. However, a significant reduction in technical difficulty can be achieved whilst retaining a large fraction of the cosmological information by performing a tomographic rather than fully 3D analysis (e.g., \citealp{Hu1999}). Here sources are separated into a number of tomographic redshift bins and we analyse the shear field averaged over these slices in redshift, i.e., we reduce the 3D shear field $\gamma(\mathbf{r})$ to a set of 2D fields $\{\gamma^{(\alpha)}(\theta,\phi)\}$, where $\alpha$ denotes the redshift bin. This collection of 2D fields will form a field vector $\field$ with covariance $\cov$ (i.e., the tomographic shear power spectra).

A forward model for the shapes of galaxies at points in 3D space necessarily requires some reference to the full 3D shear field; by restricting ourselves to a tomographic analysis without modelling the fluctuation of the field within the redshift bins, we cannot forward model a catalogue of individual galaxy shapes, angular positions and redshifts. It is possible, however, to write down a forward model for the average ellipticities of sources binned in redshift at angular positions on the sky with reference only to the tomographic shear fields. Therefore we are forced to process the catalogue of galaxy shapes and positions further into pixelized 2D maps of the average shapes in each pixel for sources in each redshift bin. This demonstrates clearly that a tomographic analysis is sub-optimal on two counts: information is lost in compressing the data from a full 3D catalogue of galaxy shapes to a collection of averages (since the detailed redshift dependence of the power spectrum contains cosmological information), and secondly in a tomographic analysis it is not possible to include the full interdependence of the shear field and galaxy redshifts shown in Fig. \ref{fig:uber_model}. The formalism developed in this work can be straightforwardly extended to perform a fully 3D shear analysis (albeit at additional computational cost) and we will investigate this in future work. 

The technical details of the processed data vector $\data$, field $\field$, noise and signal covariances $\noise$ and $\cov$ for a tomographic shear analysis are described in detail in \S \ref{sec:data_vector} and \ref{sec:shear_power} below.
\subsubsection{The data vector}
\label{sec:data_vector}
Lensing can be observed through a change in observed galaxy ellipticities. In the weak lensing limit, observed (complex) galaxy shapes $\epsilon$ are modified from their intrinsic unlensed values $\epsilon_0$ by the complex shear field $\gamma = \gamma_1 + i\gamma_2$, according to,
\begin{align}
\epsilon = \epsilon_0 + \gamma.
\end{align}
Under the assumption that $\langle\epsilon_0\rangle=0$, the observed ellipticities provide a simple unbiased point estimator for the shear. We would like to build estimated (noisy) pixelized maps of the shear field in a number of tomographic redshift bins from a catalogue of galaxy shapes, angular positions and redshifts; we can estimate the shear in a pixel $p$ averaged over redshift bin $\alpha$ by averaging the galaxy ellipticities in that pixel,
\begin{align}
\hat{\gamma}^{(\alpha)}_p = \f{1}{N^{(\alpha)}_p}\sum_{g\;\mathrm{in}\;\mathrm{pixel}\;p,\;\mathrm{bin}\;\alpha}\epsilon_g,
\end{align} 
where $N^{(\alpha)}_p$ is the number of sources in pixel $p$ and redshift bin $\alpha$. This provides an unbiased estimator for the tomographic shear field, i.e., the 3D shear $\gamma(\theta, \phi, z)$ averaged over the pixel $p$ and the galaxy redshift distribution for galaxies in redshift bin $\alpha$,
\begin{align}
\label{tomographic_field_integral}
\gamma^{(\alpha)}_p = \f{1}{\Delta\Omega_p}\int_{\mathrm{pix}\;p}d\Omega\int \gamma(\theta, \phi, z)p_{(\alpha)}(z)dz,
\end{align}
where $\Omega$ denotes solid angle, $\Delta\Omega_p$ is the solid angle of pixel $p$, $\gamma(\theta, \phi, z)$ is the full 3D shear field and $p_{(\alpha)}(z)dz$ is the redshift distribution for sources in redshift bin $\alpha$ (normalized to one over the bin). We hence assume a linear model for $\hat{\gamma}^{(\alpha)}_p$,
\begin{align}
\hat{\gamma}^{(\alpha)}_p = \gamma^{(\alpha)}_p + \epsilon^{(\alpha)}_p,\;\epsilon^{(\alpha)}_p\sim\mathcal{N}\left(0, \frac{\sqrt{2}\sigma_\epsilon}{\sqrt{N^{(\alpha)}_p}}\right)
\end{align}
where $\sigma_\epsilon^2$ is the variance of intrinsic galaxy ellipticities (per component), $\mathcal{N}(\mu, \sigma)$ denotes the Gaussian distribution with mean $\mu$ and variance $\sigma^2$ and we are assuming that $\epsilon^{(\alpha)}_p$ are uncorrelated between pixels (although this assumption can be straightforwardly lifted). Note that even if the ellipticity distribution is not Gaussian, provided many sources contribute to each pixel average the noise will become Gaussian according to the central limit theorem.

The tomographic shear fields are complex with two real components, $\gamma_p^{(\alpha)} = \gamma_{1,p}^{(\alpha)} + i\gamma_{2,p}^{(\alpha)}$; we take our data vector $\data$ to be composed of the two components of the complex shear for every pixel and redshift bin:
\begin{align}
\data = (\hat\gamma_{1,p=1}^{(1)}, &\hat\gamma_{2,p=1}^{(1)}, \hat\gamma_{1,p=1}^{(2)}, \hat\gamma_{2,p=1}^{(2)}, \dots, \nn \\
&\dots, \hat\gamma_{1,p=2}^{(1)},\hat\gamma_{2,p=2}^{(1)}, \hat\gamma_{1,p=2}^{(2)}, \hat\gamma_{2,p=2}^{(2)}, \dots).
\end{align}
Under the assumptions described above, the data vector is described by a linear model $\data = \field + \mathbf{n}$ where the field $\field$ is the collection of tomographic shear maps (whose statistics are described in \S \ref{sec:shear_power}) and the noise $\mathbf{n}$ has covariance,
\begin{align}
\langle\mathbf{n}\mathbf{n}^\mathrm{T}\rangle=\noise = \mathrm{diag}\Bigg( &\frac{\sigma_\epsilon^2}{N^{(1)}_{p=1}}, \frac{\sigma_\epsilon^2}{N^{(1)}_{p=1}},\frac{\sigma_\epsilon^2}{N^{(2)}_{p=1}}, \frac{\sigma_\epsilon^2}{N^{(2)}_{p=1}}, \dots  , \nn \\
& \dots,\frac{\sigma_\epsilon^2}{N^{(1)}_{p=2}}, \frac{\sigma_\epsilon^2}{N^{(1)}_{p=2}}, \frac{\sigma_\epsilon^2}{N^{(2)}_{p=2}}, \frac{\sigma_\epsilon^2}{N^{(2)}_{p=2}}, \dots\Bigg).
\end{align}
For masked pixels there are no observed sources contributing to $\hat\gamma_{p}^{(\alpha)}$, so the noise covariance for these pixels is taken to be infinite.
\subsubsection{The signal: tomographic shear fields and their covariance}
\label{sec:shear_power}
The tomographic shear fields described in \eqref{tomographic_field_integral} are two-dimensional isotropic random fields with spin-weight-2 on the angular sky. Since the fields are isotropic, their (spin-2) spherical harmonic coefficients are uncorrelated, making harmonic space a particularly convenient basis. The expansion coefficients and two-point statistics of the complex shear (split into $E$- and $B$-mode components) are given by
\begin{align}
&\gamma^{\mathrm{E}(\alpha)}_{\ell m} = \f{1}{2}\int\left[\gamma^{(\alpha)}(\bs\phi)_2Y^*_{\ell m}(\bs\phi) + \gamma^{*(\alpha)}(\bs\phi)_{-2}Y^*_{\ell m}(\bs\phi)\right]d\Omega,\nn \\
&\gamma^{\mathrm{B}(\alpha)}_{\ell m} = -\f{i}{2}\int\left[\gamma^{(\alpha)}(\bs\phi)_2Y^*_{\ell m}(\bs\phi) - \gamma^{*(\alpha)}(\bs\phi)_{-2}Y^*_{\ell m}(\bs\phi)\right]d\Omega,\nn \\
&\langle\gamma^{\mathrm{E}(\alpha)*}_{\ell m}\gamma^{\mathrm{E}(\beta)}_{\ell' m'}\rangle = C^\mathrm{EE}_{\ell,\alpha\beta}\delta_{mm'}\delta_{\ell\ell'}, \nn \\
&\langle\gamma^{\mathrm{E}(\alpha)*}_{\ell m}\gamma^{\mathrm{B}(\beta)}_{\ell' m'}\rangle = C^\mathrm{EB}_{\ell,\alpha\beta}\delta_{mm'}\delta_{\ell\ell'}, \nn \\
&\langle\gamma^{\mathrm{B}(\alpha)*}_{\ell m}\gamma^{\mathrm{B}(\beta)}_{\ell' m'}\rangle = C^\mathrm{BB}_{\ell,\alpha\beta}\delta_{mm'},\delta_{\ell\ell'},
\end{align}
where $_{\pm 2}Y_{\ell m}$ are the spin-weight $\pm 2$ spherical harmonics, $C^\mathrm{EE}_{\ell,\alpha\beta}$, $C^\mathrm{EB}_{\ell,\alpha\beta}$ and $C^\mathrm{EE}_{\ell,\alpha\beta}$ are the $E$-mode, $B$-mode and cross $EB$ angular power spectra between tomographic bins $\alpha$ and $\beta$ and $\delta_{nm}$ is the Kronecker-delta. Typically, cosmological models predict negligible $B$-mode, so $C^\mathrm{BB}_{\ell,\alpha\beta} \simeq 0$, and parity considerations require $C^\mathrm{EB}_{\ell,\alpha\beta} \simeq 0$. However, systematic effects could give rise to non-zero $B$-modes, so in a weak lensing analysis the estimation of the $B$-mode power is nonetheless useful as it provides a test for systematic effects.

In the Limber approximation \citep{Limber1954}, the $E$-mode tomographic shear power spectra are given by \citep{Kaiser1992, Kaiser1998, Hu1999, Hu2002a, Takada2004},
\begin{align}
\label{limber_power}
C^\mathrm{EE}_{\ell,\alpha\beta} &= \int \f{d\chi}{\chi_\mathrm{m}^2(\chi)}\;w_{(\alpha)}(\chi)w_{(\beta)}(\chi)(1+z)^2P_\delta\left(\f{\ell}{\chi_\mathrm{m}(\chi)}; \chi\right),
\end{align}
where $\chi$ is comoving distance, $P(k; \chi)$ is the 3D matter power spectrum and $\chi_\mathrm{m}(\chi)$ is the transverse comoving distance corresponding to comoving distance $\chi$. The lensing weight functions $w_{(\alpha)}(\chi)$ are given by
\begin{align}
w_{(\alpha)}(\chi)=
\f{3\Omega_\mathrm{m}H_0^2}{2}\chi_\mathrm{m}(\chi)\int_{\chi}^{\chi_\mathrm{H}} d\chi'\;n_{(\alpha)}(\chi')\f{\chi_\mathrm{m}(\chi'-\chi)}{\chi_\mathrm{m}(\chi')}
\end{align}
where $n_{(\alpha)}(\chi)d\chi = p_{(\alpha)}(z)dz$ is the redshift distribution for galaxies in redshift bin $\alpha$ (normalized to one over the bin).

The fields under consideration are the collection of tomographic shear maps $\{\gamma^{(\alpha)}(\theta,\phi)\}$. In the context of the hierarchical model in Fig. \ref{fig:vanilla_model}, the field $\field$ represented in harmonic space contains the set of harmonic coefficients $\{\gamma^{E(\alpha)}_{\ell m},\gamma^{B(\alpha)}_{\ell m}\}$ arranged into a vector:
\begin{align}
\label{field_vector}
&\field = \left(\field_{00}, \field_{1-1}, \field_{10}, \field_{11}\dots\field_{\ell m}\dots\right), \nn \\
&\field_{\ell m} = \left(\gamma^{\mathrm{E}(1)}_{\ell m}, \gamma^{\mathrm{E}(2)}_{\ell m}, \dots, \gamma^{\mathrm{E}(n_\mathrm{bins})}_{\ell m}, \gamma^{\mathrm{B}(1)}_{\ell m}, \gamma^{\mathrm{B}(2)}_{\ell m}, \dots \gamma^{\mathrm{B}(n_\mathrm{bins})}_{\ell m}\right).
\end{align}
The full covariance matrix $\cov$ of the field $\field$ will then be block-diagonal, where each $\ell m$-mode contributes one block $\cov_{\ell m}$,
\begin{align}
\langle\field\field^\dagger\rangle = \cov &= \mathrm{diag}\left(\cov_{00}, \cov_{1-1}, \cov_{10}, \cov_{11}\dots \cov_{\ell m}\dots\right), \nn \\
& = \mathrm{diag}\left(\cov_0, \cov_1,\cov_2\dots\cov_\ell\dots\right),
\end{align}
where $\cov_\ell = \cov_{\ell m}\otimes \mathbf{I}_{2\ell + 1}$ is the block diagonal contribution for a given $\ell$ mode, with $2\ell + 1$ diagonal sub-blocks for each $m$ mode at the given $\ell$,
\begin{align} 
&\cov_{\ell m}= \left( \begin{array}{ccccccc}
C^{\mathrm{EE}}_{\ell, 11} & C^{\mathrm{EE}}_{\ell, 12} & \dots & C^{\mathrm{EB}}_{\ell, 11} & C^{\mathrm{EB}}_{\ell, 12} & \dots \\
C^{\mathrm{EE}}_{\ell, 21} & C^{\mathrm{EE}}_{\ell, 22} & \dots &C^{\mathrm{EB}}_{\ell, 21} & C^{\mathrm{EB}}_{\ell, 22} & \dots \\
\vdots & \vdots &  \ddots & \vdots& \vdots & \ddots \\
C^{\mathrm{BE}}_{\ell, 11} & C^{\mathrm{BE}}_{\ell, 12} & \dots & C^{\mathrm{EB}}_{\ell, 11} & C^{\mathrm{BB}}_{\ell, 12} & \dots \\
C^{\mathrm{BE}}_{\ell, 21} & C^{\mathrm{BE}}_{\ell, 22} & \dots &C^{\mathrm{BB}}_{\ell, 21} & C^{\mathrm{BB}}_{\ell, 22} & \dots \\
\vdots & \vdots &  \ddots & \vdots& \vdots & \ddots \end{array} \right).
\end{align}
$\mathbf{I}_n$ is the $n\times n$ identity matrix, $\otimes$ is the Kronecker product and again $C_{\ell,\alpha\beta}$ are the tomographic angular power spectra between redshift bins $\alpha$ and $\beta$. Note that, in principle, the shear covariance could contain contributions from both cosmic shear (as described above) and also intrinsic alignments; see \S \ref{sec:ia} for a discussion of including intrinsic alignments into this framework.

\subsubsection{Flat sky approximation}
\label{sec:flat}
In the limit where we have a small survey area, we can make the flat sky-approximation and replace spherical harmonic transforms with Fourier transforms. In this limit, the $E$- and $B$-mode shear coefficients and power spectra are given by:
\begin{align}
&\gamma^{\mathrm{E}(\alpha)}_{\bs{\ell}} = \f{1}{2}\cdot\f{1}{2\pi}\int\left[\gamma^{(\alpha)}(\bs{\phi})\varphi_{\bs{\ell}}^*e^{-i\bs{\ell}\cdot\bs{\phi}} + \gamma^{*(\alpha)}(\bs{\phi})\varphi_{\bs{\ell}}e^{-i\bs{\ell}\cdot\bs{\phi}}\right]d\Omega,\nn \\
&\gamma^{\mathrm{B}(\alpha)}_{\bs{\ell}} = -\f{i}{2}\cdot\f{1}{2\pi}\int\left[\gamma^{(\alpha)}(\bs{\phi})\varphi_{\bs{\ell}}^*e^{-i\bs{\ell}\cdot\bs{\phi}} - \gamma^{*(\alpha)}(\bs{\phi})\varphi_{\bs{\ell}}e^{-i\bs{\ell}\cdot\bs{\phi}}\right]d\Omega,\nn \\
&\langle\gamma^{\mathrm{E}(\alpha)*}_{\bs{\ell}}\gamma^{\mathrm{E}(\beta)}_{\bs{\ell}'}\rangle = C^\mathrm{EE}_{\ell,\alpha\beta}\delta_{{\bs{\ell}}{\bs{\ell}}'}, \nn \\
&\langle\gamma^{\mathrm{E}(\alpha)*}_{\bs{\ell}}\gamma^{\mathrm{B}(\beta)}_{\bs{\ell}'}\rangle = C^\mathrm{EB}_{\ell,\alpha\beta}\delta_{{\bs{\ell}}{\bs{\ell}}'}, \nn \\
&\langle\gamma^{\mathrm{B}(\alpha)*}_{\bs{\ell}}\gamma^{\mathrm{B}(\beta)}_{\bs{\ell}'}\rangle = C^\mathrm{BB}_{\ell,\alpha\beta}\delta_{{\bs{\ell}}{\bs{\ell}}'},
\end{align}
where $\bs\ell = (\ell_x, \ell_y)$, the phase factor $\varphi_{\bs{\ell}} = - (\ell_x^2 - \ell_y^2 + 2i\ell_x\ell_y)/\ell^2$ and the angular power spectra are as previously. The field vector for the flat-sky approximation shear coefficients is then given by
\begin{align}
&\field = \left(\field_{\bs{\ell}_1}, \field_{\bs{\ell}_2}, \field_{\bs{\ell}_3}, \dots\field_{\bs{\ell}_i}\dots\right), \nn \\
&\field_{\bs{\ell}} = \left(\gamma^{\mathrm{E}(1)}_{\bs{\ell}}, \gamma^{\mathrm{E}(2)}_{\bs{\ell}}, \dots, \gamma^{\mathrm{E}(n_\mathrm{bins})}_{\bs{\ell}}, \gamma^{\mathrm{B}(1)}_{\bs{\ell}}, \gamma^{\mathrm{B}(2)}_{\bs{\ell}}, \dots \gamma^{\mathrm{B}(n_\mathrm{bins})}_{\bs{\ell}}\right).
\end{align}
The full covariance matrix $\cov$ of the field $\field$ will again be block-diagonal, with each ${\bs{\ell}}$-mode contributing one block,
\begin{align}
&\langle\field\field^\dagger\rangle = \cov = \mathrm{diag}\left(\cov_{\bs{\ell}_1}, \cov_{\bs{\ell}_2}, \cov_{\bs{\ell}_3}, \cov_{\bs{\ell}_4}\dots \cov_{\bs{\ell}_i}\dots\right), \nn \\
& = \mathrm{diag}\left(\cov_0, \cov_1,\cov_2\dots\cov_\ell\dots\right),
\end{align}
where here $\cov_\ell = \cov_{\ell_x \ell_y}\otimes \mathbf{I}_{n}$ is the block diagonal contribution from each $\ell$ mode, and $\cov_{\ell_x \ell_y}$ are the $n$ sub-blocks for each of the $\bs{\ell} = (\ell_x, \ell_y)$ modes with $|\bs{\ell}| = \ell$,
\begin{align}
&\cov_{\ell_x \ell_y}= \left( \begin{array}{ccccccc}
C^{\mathrm{EE}}_{\ell, 11} & C^{\mathrm{EE}}_{\ell, 12} & \dots & C^{\mathrm{EB}}_{\ell, 11} & C^{\mathrm{EB}}_{\ell, 12} & \dots \\
C^{\mathrm{EE}}_{\ell, 21} & C^{\mathrm{EE}}_{\ell, 22} & \dots &C^{\mathrm{EB}}_{\ell, 21} & C^{\mathrm{EB}}_{\ell, 22} & \dots \\
\vdots & \vdots &  \ddots & \vdots& \vdots & \ddots \\
C^{\mathrm{BE}}_{\ell, 11} & C^{\mathrm{BE}}_{\ell, 12} & \dots & C^{\mathrm{EB}}_{\ell, 11} & C^{\mathrm{BB}}_{\ell, 12} & \dots \\
C^{\mathrm{BE}}_{\ell, 21} & C^{\mathrm{BE}}_{\ell, 22} & \dots &C^{\mathrm{BB}}_{\ell, 21} & C^{\mathrm{BB}}_{\ell, 22} & \dots \\
\vdots & \vdots &  \ddots & \vdots& \vdots & \ddots \end{array} \right).
\end{align}

\section{Hierarchical tomographic shear map-power spectrum inference}
\label{sec:map_power_inference}
In the following section we develop the machinery for jointly inferring the tomographic shear fields $\field$ and covariance (power spectra) $\cov$ as described in \S \ref{sec:shear_power}, from observed noisy tomographic shear maps $\data$ described in \S \ref{sec:data_vector}. However, the formalism developed here applies to the more general problem of jointly inferring any Gaussian field and its covariance given a noisy estimate of the field and the data is described by a Gaussian linear model, i.e., $\data = \field + \mathbf{n}$, where $\mathbf{n}$ is Gaussian noise of known covariance $\noise$. The methods described in this section are similar to approaches taken to power spectrum inference for CMB temperature \citep{Wandelt2004, Eriksen2004, Odwyer2004, Chu2005, Komatsu2011, Planck2013XV}, CMB polarization \citep{Larson2007, Eriksen2007, Komatsu2011, Karakci2013} and large-scale structure \citep{Jasche2010, Jasche2012, Jasche2013}, but here specialized for application to weak lensing.
\subsection{Vanilla map-power spectrum inference and the Wiener filter}
\label{sec:vanilla}
The basic forward model for the tomographic shear maps is described in Fig. \ref{fig:vanilla_model} and is understood as follows: We begin by fixing a power spectrum (chosen from some prior). This power spectrum generates a shear field via the density $P(\field|\cov)$ (which is Gaussian under the current assumptions). We then specify a noise covariance matrix $\noise$ and add noise to the realised shear map to give a realisation of the data, via the conditional density $P(\data|\field, \noise)$. The graph in Fig. \ref{fig:vanilla_model} shows us explicitly the conditional structure of the full posterior $P(\cov, \field|\data)$, since we can simply write the posterior as the product of the conditional densities (and priors) appearing on the edges of the graph:
\begin{align}
\label{posterior_vanilla}
P(\cov, \field|\data, \noise) = \f{P(\data|\field , \noise)P(\field |\cov)P(\cov)}{P(\data)},
\end{align}
where the conditional densities are given by
\begin{align}
&P(\data|\field , \noise) = \frac{1}{\sqrt{(2\pi)^N|\noise|}}e^{-\frac{1}{2}(\mathbf{d}-\mathbf{s} )^\dagger \noise^{-1}(\mathbf{d}-\mathbf{s} )},\nn \\
&P(\mathbf{s} |\cov) =  \frac{1}{\sqrt{(2\pi)^N|\cov|}}e^{-\frac{1}{2}\mathbf{s} ^\dagger \cov^{-1}\mathbf{s}},
\end{align}
and $N = 2\times n_\mathrm{bins}\times n_\mathrm{pix}$ is the length of the vectors $\data$ and $\field$ for $n_\mathrm{bins}$ tomographic shear maps each containing $n_\mathrm{pix}$ pixels (and the factor of $2$ is due to including both $E$- and $B$-mode degrees-of-freedom). For the prior on the covariance matrix we take a Jeffreys prior \citep{Jeffreys1961} on each sub-block of the covariance,
\begin{align}
P(\cov) \propto \prod_\ell|\cov_\ell|^{-(p+1)/2},
\end{align}
where the sub-blocks $\cov_\ell$ are $p\times p$ matrices with $p = 2\times n_\mathrm{bins}$. For a discussion of non-informative priors for covariance matrix inference, see \citet{Daniels1999} and references therein.

Hierarchical models lend themselves naturally to Gibbs or block-MCMC sampling, where at each step we draw a sample of each parameter in turn conditional on all others. For the current model, this means iteratively drawing samples of the shear field and covariance matrix:
\begin{align}
&\cov^{i+1} \leftarrow P(\cov|\field^i) \nn \\
&\field^{i+1} \leftarrow P(\field|\cov^i, \data, \noise).
\end{align}
In order to build a Gibbs or block-MCMC sampler, then, we must know the conditional densities for each parameter given all others. The graphical model makes writing down these conditionals particularly straightforward; the conditional density for any parameter is given by the product of the conditional densities on all incoming and outgoing edges in the graph (upto a normalization constant). The conditional density of the shear map is hence:
\begin{align}
\label{map_conditional_vanilla}
P(\field|\cov, \noise, \mathbf{d}) &\propto P(\data|\field, \noise)P(\field|\cov) \nn \\
&= \frac{1}{\sqrt{(2\pi)^N|\cov_\mathrm{WF}|}}e^{-\frac{1}{2}(\field-\data_\mathrm{WF} )^\dagger \cov_\mathrm{WF}^{-1}(\field-\data_\mathrm{WF} )},
\end{align}
where $\data_\mathrm{WF}=(\cov^{-1}+\noise^{-1})^{-1}\noise^{-1}\data$ is the Wiener filter of the data and the covariance $\mathbf{C}_\mathrm{WF}=(\cov^{-1}+\noise^{-1})^{-1}$. Similarly the conditional density of the signal covariance is given by:
\begin{align}
\label{power_conditional_vanilla}
P(\cov|\mathbf{s})&= \f{P(\mathbf{s}|\cov)P(\cov)}{P(\mathbf{s})}.
\end{align}
Since the covariance matrix is block diagonal, with each $\ell m$-mode contributing one block, and due to isotropy every $m$-mode for a given $\ell$ has the same power, we can factorize the conditional on the covariance $\cov$ into conditional distributions on each $\ell$-mode covariance $\cov_\ell$,
\begin{align}
\label{power_mode_conditional_vanilla}
P(\cov|\mathbf{s}) &= \prod_\ell P(\cov_\ell|\mathbf{s}), \nn \\
P(\cov_\ell|\mathbf{s})&=\f{|\bs{\Gamma}_\ell|^{(2\ell + 1)/2}|\cov_\ell|^{-(2\ell+2+ p)/2}}{2^{(2\ell + 1)p/2}\Gamma_p(\ell+\f{1}{2})}\exp\left[-\f{1}{2}\mathrm{tr}(\cov_\ell^{-1}\boldsymbol{\Gamma_\ell})\right] \nn \\
&=\mathcal{W}^{-1}(\bs{\Gamma}_\ell, \nu_\ell),
\end{align}
where $\bs{\Gamma}_\ell =\sum_m \field_{\ell m}\field_{\ell m}^\dagger$ and $\Gamma_p(\cdot)$ is the multivariate gamma function. In the second line of Eq. \eqref{power_mode_conditional_vanilla} we recognise the conditional density of the $\cov_\ell$ to be the inverse-Wishart distribution with support $\bs{\Gamma}_\ell$ and $\nu_\ell= 2\ell+1$ degrees-of-freedom, denoted by $\mathcal{W}^{-1}(\bs{\Gamma}_\ell, \nu_\ell)$.
\subsubsection*{Map sampling}
Sampling from the map conditional on the signal covariance involves generating a Gaussian random vector whose mean is the Wiener filter of the data $\data_\mathrm{WF}=(\cov^{-1}+\noise^{-1})^{-1}\noise^{-1}\data$ and with covariance $\mathbf{C}_\mathrm{WF}=(\cov^{-1}+\noise^{-1})^{-1}$. Whilst this is simple in principle, computing the Wiener filter involves inverting the matrix $(\cov^{-1}+\noise^{-1})$ which has size $N\times N$ where $N=2\times n_\mathrm{bins}\times n_\mathrm{pix}$ for $n_\mathrm{bins}$ tomographic shear maps containing $n_\mathrm{pix}$ pixels (and the factor of $2$ is due to including both $E$- and $B$-mode degrees-of-freedom). Since in applications the number of pixels is typically $n_\mathrm{pix}\sim 10^4-10^6$ depending on the survey size and the angular scales of interest, this numerical matrix inversion is not computationally feasible by brute force methods. Progress can be made if there exists a basis where both $\cov$ and $\noise$ are sparse. For cosmic shear, the signal covariance is sparse in harmonic space whilst the noise covariance is sparse in pixel-space (for weakly or uncorrelated pixel noise). In the idealised case where the pixel noise is both homogeneous and isotropic, $\noise$ will be proportional to the identity matrix and is hence sparse (proportional to the identity) in any basis. The matrix inversion can then be performed in harmonic space where both the signal covariance and noise are now sparse. However, in practice the pixel noise will not be homogeneous and isotropic and it will not in general be possible to find a single basis where both $\cov$ and $\noise$ are sparse. For example, if the pixel noise is uncorrelated but the number of sources contributing to each pixel varies (as is inevitable in practice), $\noise$ is no longer isotropic; it is diagonal in pixel space but will \emph{not} be sparse in harmonic space. Pixel-pixel noise correlations would also result in $\noise$ and $\cov$ not being sparse in the same basis. In these cases, numerical implementations of the Wiener filter have traditionally relied on Krylov space methods, such as conjugate gradients, to solve the high-dimensional systems of equations (see e.g., \citealp{Kitaura2008} and references therein). Recently a particularly elegant approach to solving the Wiener filter equation was proposed, where an additional messenger field is introduced to mediate between two different bases in which the signal and noise covariances are respectively sparse, bypassing the issue of directly inverting the high-dimensional matrices \citep{Elsner2012, Elsner2013, Jasche2015}. We adopt this approach in \S \ref{sec:messenger} and apply it to simulated data in \S \ref{sec:results}.
\subsubsection*{Power spectrum sampling}
In order to draw samples of the signal covariance for fixed shear field we simply need to generate inverse-Wishart distributed random matrices with $\nu = 2\ell+1$ degrees-of-freedom and support $\boldsymbol{\Gamma}_\ell$. Drawing samples from the inverse-Wishart distribution $\mathcal{W}^{-1}(\boldsymbol{\Gamma}, \nu)$ can be straightforwardly performed as follows:
\begin{enumerate}
\item Generate $\nu$ Gaussian random vectors $\mathbf{x}_i\sim\mathcal{N}(\mathbf{0}, \boldsymbol{\Gamma}^{-1})$.
\item Construct the sum of outer products of the vectors $\{\mathbf{x}_i\}$, i.e. $\mathbf{X} = \sum_{i=1}^\nu \mathbf{x}_i\mathbf{x}_i^\mathrm{T}$.
\item Take the inverse of $\mathbf{X}$, then $\mathbf{X}^{-1}\sim\mathcal{W}^{-1}(\boldsymbol{\Gamma}, \nu)$ as required.
\end{enumerate}
\subsubsection*{Survey Mask}
One of the major benefits of jointly inferring the shear map and power spectrum in a hierarchical setting is the ease with which masked regions can be accounted for. Masked pixels are treated as regions where the data provides no information, i.e., the noise covariance for masked pixels is taken to be infinite. However, this does not mean that we are totally ignorant about the field in those pixels. When generating samples of the field for a fixed covariance, $\field^{i+1}\leftarrow P(\field|\cov^i, \data, \noise) \propto P(\data|\field, \noise)P(\field|\cov^i)$, there are two contributions to the density $P(\field|\cov^i, \data, \noise)$; the data provides information about the field through the likelihood $P(\data|\field, \noise)$ and the covariance provides additional information via $P(\field|\cov^i)$. For pixels in which the noise covariance is infinite, all of the information comes from the prior on the field $P(\field|\cov^i)$ at that sampling step, but nonetheless since the covariance is also being explored we are able to make inferences about the field inside the masked regions (albeit at lower signal-to-noise).
\subsection{Messenger field}
\label{sec:messenger}
Sampling from the posterior distribution of the shear map and covariance described in \S \ref{sec:vanilla} is conceptually simple and and can be reduced to drawing (multivariate) Gaussian and inverse-Wishart random variates. However, drawing samples of the map conditional on the signal covariance involves computing the Wiener filter of the data and inverting the matrix $(\cov^{-1} + \noise^{-1})$, which has size $\sim n_\mathrm{pix}\times n_\mathrm{pix}$. If it is not possible to find a basis in which $\cov$ and $\noise$ are both sparse simultaneously (such as in the  realistic case of anisotropic noise as discussed in \S \ref{sec:vanilla}), this matrix inversion results in a formidable computational bottleneck. Fortunately, direct inversion of $(\cov^{-1} + \noise^{-1})$ can be avoided by introducing an auxiliary Gaussian distributed messenger field $\messenger$ that mediates between the bases in which $\cov$ and $\noise$ are respectively sparse. This elegant idea was introduced by \citet{Elsner2012, Elsner2013} and further developed by \citet{Jasche2015}; the approach taken here is close to \citet{Jasche2015} but generalized to deal with non-diagonal signal covariance matrices. 
\begin{figure}
\begin{center}
\includegraphics{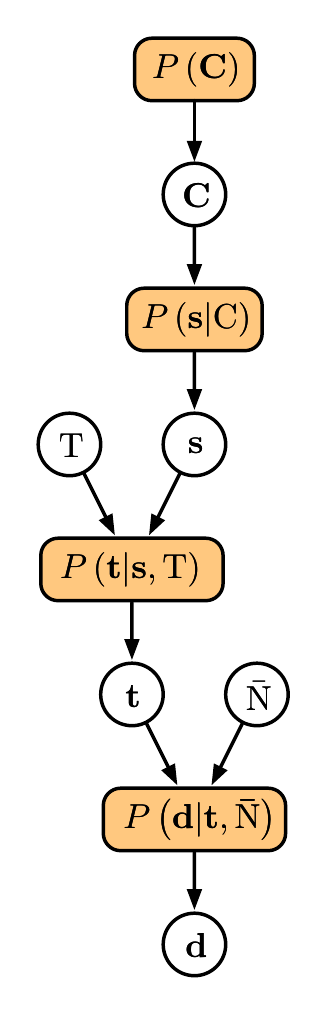}
\end{center}
\caption{Hierarchical forward model for shear map-power spectrum inference with the addition of a messenger field: the shear power spectrum is drawn from some prior distribution, a realization of the shear field is then generated given the power spectrum, isotropic noise with covariance $\isotropic$ is added to give a realization of the messenger field $\messenger$ and finally an anisotropic noise component is added with covariance $\aniso$ to realize a noisy shear map $\data$.}
\label{fig:messenger_model}
\end{figure}

The inclusion of the messenger field effectively allows us to split the noise $\noise$ into two parts that are dealt with separately: an isotropic part $\isotropic=\tau\mathbf{I}$ where $\tau \leq \mathrm{min}\left[\mathrm{diag}(\noise)\right]$, and the remaining anisotropic noise $\aniso = \noise - \isotropic$. The generative model for the data via the signal covariance $\cov$, shear field $\field$ and messenger field $\messenger$ is summarised in Fig. \ref{fig:messenger_model} and is understood as follows: the signal covariance $\cov$ generates a shear map, the shear map plus isotropic noise drawn from $\mathcal{N}(\mathbf{0}, \isotropic)$ generates a realisation of the messenger field $\messenger$, and finally the messenger field plus an additional anisotropic noise component drawn from $\mathcal{N}(\mathbf{0}, \aniso)$ generates a realisation of the data $\data$. Importantly, note that the introduction of an additional level in the hierarchy separates the signal covariance $\cov$ from the anisotropic noise covariance $\aniso$, connecting them only via $\isotropic\propto\mathbf{I}$ which is diagonal in any basis; this is the essential function of the messenger field.  

The posterior and conditional distributions are:
\begin{align}
&P(\cov, \field, \messenger|\data,\isotropic, \aniso) = \f{P(\data|\messenger, \aniso)P(\messenger|\field, \isotropic)P(\field |\cov)P(\cov)}{P(\data)}, \nn \\
&P(\data|\messenger, \aniso) =\frac{1}{\sqrt{(2\pi)^N|\aniso|}}e^{-\frac{1}{2}(\mathbf{d}-\messenger )^\dagger \aniso^{-1}(\mathbf{d}-\messenger )}, \nn \\
&P(\field |\cov) = \frac{1}{\sqrt{(2\pi)^N|\cov|}}e^{-\frac{1}{2}\field^\dagger \cov^{-1}\field }, \nn \\
&P(\messenger|\field, \isotropic) = \frac{1}{\sqrt{(2\pi)^N|\isotropic|}}e^{-\frac{1}{2}(\messenger-\field)^\dagger \isotropic^{-1}(\messenger -\field)}.
\end{align}
Note that marginalising the posterior $P(\cov, \field, \messenger|\data)$ over the messenger field $\messenger$ recovers the joint posterior distribution for the shear field and signal covariance $P(\cov, \field|\data)$ of Eq. \eqref{posterior_vanilla}, so sampling from the posterior $P(\cov, \field, \messenger|\data)$ and marginalizing over $\field$ and $\messenger$ is completely equivalent to sampling from $P(\cov, \field|\data)$ and marginalizing over $\field$; both methods will generate samples from the marginal posterior $P(\cov|\data)$ as desired.

In order to block-MCMC or Gibbs sample from the posterior $P(\cov, \field, \messenger|\data)$ we need to iteratively draw samples from $\cov$, $\field$ and $\messenger$ conditional on all other parameters,
\begin{align}
&\cov^{i+1} \leftarrow P(\cov|\field^i), \nn \\
&\field^{i+1} \leftarrow P(\field|\messenger^i, \cov^i, \isotropic), \nn \\
&\messenger^{i+1} \leftarrow P(\messenger|\field^i, \data, \aniso),
\end{align}
where the conditional densities (from the graph in Fig. \ref{fig:messenger_model}) are given by
\begin{align}
P(\field|\cov, \isotropic, \messenger) &\propto P(\messenger|\field, \isotropic)P(\field|\cov) \nn \\
&= \frac{1}{\sqrt{(2\pi)^N|\mathbf{Q}_\field|}}e^{-\frac{1}{2}(\field - \bs{\mu}_\field )^\dagger \mathbf{Q}_\field^{-1}(\field - \bs{\mu}_\field )}, \nn \\
P(\messenger|\field, \isotropic, \aniso, \data) &\propto P(\data|\messenger, \aniso)P(\messenger|\isotropic, \field) \nn \\
&= \frac{1}{\sqrt{(2\pi)^N|\mathbf{Q}_\messenger|}}e^{-\frac{1}{2}(\messenger - \bs{\mu}_\messenger)^\dagger \mathbf{Q}_\messenger^{-1}(\messenger - \bs{\mu}_\messenger)}, \nn \\
P(\cov_\ell|\mathbf{s})&=\mathcal{W}^{-1}(\bs{\Gamma}_\ell, \nu_\ell),
\end{align}
and the (conditional) shear field and messenger field means and covariances are given by,
\begin{align}
\bs{\mu}_\field &= (\cov^{-1} + \isotropic^{-1})^{-1}\isotropic^{-1}\messenger, \nn \\
\mathbf{Q}_\field & = (\cov^{-1} + \isotropic^{-1})^{-1}, \nn \\
\bs{\mu}_\messenger &= (\isotropic^{-1} + \aniso^{-1})^{-1}\isotropic^{-1}\field + (\isotropic^{-1} + \aniso^{-1})^{-1}\aniso^{-1}\data, \nn \\
\mathbf{Q}_\messenger & = (\isotropic^{-1} + \aniso^{-1})^{-1}.
\end{align}
Crucially, note that the introduction of the messenger field has isolated the signal covariance $\cov$ from the anisotropic noise $\aniso$, where $\cov$ and $\aniso$ now only appear in combination with the isotropic noise component $\isotropic$ which is diagonal in any basis, since $\isotropic = \tau\mathbf{I}$. Thus all of the necessary matrix inversions can now be performed in bases where the matrices are sparse, eliminating the need to solve the Wiener filter equation for high-dimensional dense matrices.
\subsubsection*{Map sampling}
Sampling from the map conditional on the signal covariance and messenger field involves generating a Gaussian random vector with mean and covariance $\bs{\mu}_\field = (\cov^{-1} + \isotropic^{-1})^{-1}\isotropic^{-1}\messenger$ and $\mathbf{Q}_\field = (\cov^{-1} + \isotropic^{-1})^{-1}$. The noise component appearing here is purely isotropic, $\isotropic\propto\mathbf{I}$, and hence $\isotropic$ is diagonal in any orthogonal basis; the matrix $\mathbf{Q}_\field = (\cov^{-1} + \isotropic^{-1})^{-1}$ is block-diagonal in harmonic space, with sub-blocks of size $p=2\times n_\mathrm{bins}$, hence the largest matrix that has to be inverted during the map-sampling step is $p\times p$.
\subsubsection*{Messenger field sampling}
Sampling from the messenger field conditional on the shear field and the data involves generating a Gaussian random vector with mean and covariance $\bs{\mu}_\messenger = (\isotropic^{-1} + \aniso^{-1})^{-1}\isotropic^{-1}\field + (\isotropic^{-1} + \aniso^{-1})^{-1}\aniso^{-1}\data$ and $\mathbf{Q}_\messenger = (\isotropic^{-1} + \aniso^{-1})^{-1}$. If the pixel noise is weakly correlated, the anisotropic noise covariance $\aniso$ will be sparse in pixel-space, and in the limit where the noise can be assumed to be completely uncorrelated $\aniso$ will be diagonal in the pixel-basis. Since $\isotropic$ is diagonal in any basis by construction, inverting the matrix $(\isotropic^{-1}+\aniso^{-1})$ is now trivial in the pixel-basis. Miraculously, by splitting the noise into an isotropic and anisotropic component and introducing a messenger field to mediate between the data and the shear field, we have created an environment where $\cov$ and $\aniso$ can be represented in bases in which they are respectively sparse simultaneously, eliminating the need to numerically invert $\sim n_\mathrm{pix}\times n_\mathrm{pix}$ matrices at each sampling step.
\subsubsection*{Power spectrum sampling}
The conditional distribution for the signal covariance is unaffected by the introduction of the messenger field into the hierarchy, so the power spectrum sampling step is again achieved by drawing inverse-Wishart random variates $\mathbf{C}_\ell^{i+1}\sim\mathcal{W}^{-1}(\boldsymbol{\Gamma}_\ell^i, \nu_\ell)$ as described in \S \ref{sec:vanilla}.
\begin{algorithm}
\caption{Joint shear field-messenger field-power spectrum Gibbs sampling. Here we make harmonic space and real space explicit, with harmonic space variables being denoted with a tilde. $\mathrm{F}[\cdot]$ represents the orthogonal transformation from real space to harmonic $E$- and $B$-mode coefficients, $\mathcal{N}(\bs\mu, \bs\Sigma)$ represents drawing from a multivariate Gaussian with mean $\bs\mu$ and covariance $\bs\Sigma$, and $\mathcal{W}^{-1}\left(\bs{\Gamma}, \nu\right)$ represents drawing from an Inverse-Wishart distribution with support $\bs\Gamma$ and $\nu$ degrees-of-freedom as described in \S \ref{sec:map_power_inference}. The organization of the vectors $\field_{\ell m}$ and $\field$ are described by Eq. \eqref{field_vector}.}
\begin{algorithmic}
\State{Initialize parameter values:}
\State{\emph{Draw $\cov$ from some prior and initialize $\field$:}}
\State{$\cov\sim P(\cov)$}
\State{$\tilde\field\sim\mathcal{N}(\bs{0}, \cov)$}
\State{$\field = \mathrm{F}^{-1}\left[\tilde\field\right]$}\vspace{1.5mm}
\State{$N$ Gibbs iterations:}
\For{$i = 0 \leftarrow N$}
\State{Sample messenger field:}
\State{\emph{Mean and Covariance:}}
\State{$\bs{\mu}_\messenger = (\isotropic^{-1} + \aniso^{-1})^{-1}\isotropic^{-1}\field + (\isotropic^{-1} + \aniso^{-1})^{-1}\aniso^{-1}\data$} 
\State{$\mathbf{Q}_\messenger  = (\isotropic^{-1} + \aniso^{-1})^{-1}$} \vspace{1.5mm}
\State{\emph{Draw Gaussian random variate:}}
\State{$\messenger \sim \mathcal{N}(\bs{\mu}_\messenger, \mathbf{Q}_\messenger)$}\vspace{1.5mm}
\State{\emph{Transform to harmonic space:}}
\State{$\tilde{\messenger} = \mathrm{F}\left[\messenger\right]$} \\
\State{Sample shear field:}
\For{all $\ell,\;m$ modes}\vspace{1.5mm}
\State{\emph{Mean and Covariance:}}
\State{$\bs{\mu}_{\field,\ell m} = (\cov_\ell^{-1} + \isotropic_\ell^{-1})^{-1}\isotropic_\ell^{-1}\tilde{\messenger}_{\ell m}$}
\State{$\mathbf{Q}_{\field,\ell} = (\cov_\ell^{-1} + \isotropic_\ell^{-1})^{-1}$}\vspace{1.5mm}
\State{\emph{Draw Gaussian random variate:}}
\State{$\tilde\field_{\ell m} \sim \mathcal{N}(\bs{\mu}_{\field,\ell m}, \mathbf{Q}_{\field,\ell})$}\vspace{1.5mm}
\EndFor
\State{\emph{Transform to real space:}}
\State{$\field = \mathrm{F}^{-1}\left[\tilde\field\right]$} \\
\State{Sample covariance matrix:}
\For{all $\ell$ modes}\vspace{1.5mm}
\State{\emph{Compute $\bs\Gamma_\ell$:}}
\State{$\bs{\Gamma}_\ell = \sum_{m=-\ell}^\ell \tilde\field_{\ell m}\tilde\field_{\ell m}^\dagger$}\vspace{1.5mm}
\State{\emph{Draw Inverse-Wishart random variate:}}
\State{$\cov_\ell \sim \mathcal{W}^{-1}\left(\bs{\Gamma}_\ell, \nu_\ell\right)$}
\EndFor
\EndFor
\end{algorithmic}
\label{algorithm:algo}
\end{algorithm}
\subsection{Cosmological parameter inference from the power spectrum and the Blackwell-Rao estimator}
\label{sec:br}
The approach described in \S \ref{sec:messenger} allows us to jointly sample from the posterior $P(\cov, \field, \messenger|\data)$. Marginalizing over $\field$ and $\messenger$ we obtain samples from the marginal posterior distribution of the power spectrum (signal covariance) $\cov$ given the data, i.e.,
\begin{align}
P(\cov|\data) = \int P(\cov, \field, \messenger|\data)d\field d\messenger.
\end{align}
Ultimately we are not interested in the power spectrum itself, but rather what it can tell us about cosmological models and parameters. Cosmological models provide a deterministic mapping between cosmological parameters and the shear power spectrum, i.e., for a given set of cosmological parameters $\bs{\theta}$ and model $M$ we can compute $\cov(\bs{\theta}, M)$. This means that if we have access to the posterior $P(\cov|\data)$, we can straightforwardly sample the posterior distribution for $\bs{\theta}$ in a given model $M$ by drawing samples from 
\begin{align}
\label{cosmo_posterior}
\bs\theta\sim P(\cov(\bs\theta, M)|\data)\frac{P(\bs\theta)}{P(\cov(\bs\theta, M))},
\end{align}
where the prior on the covariance $\cov$ is effectively replaced by a prior on $\bs\theta$ at this stage, through the ratio $P(\bs\theta)/P(\cov(\bs\theta, M)$. In order to sample $\bs{\theta}$ we would like access to the full smooth posterior $P(\cov|\data)$, rather than just a set of samples of the map and covariance $\{\field^i,\cov^i\}$. The Blackwell-Rao estimator \citep{Gelfand1990} provides an efficient and very accurate way of estimating the smooth density $P(\cov|\data)$ from a set of samples of the field $\{\field\} = \{\field^1, \dots, \field^n\}$.

The Blackwell-Rao estimator can be understood as follows. When block-MCMC or Gibbs sampling the field and signal covariance, it is perfectly allowed to draw many samples of $\cov$ for each field sample. In the limit where we take a large number of covariance samples at every step, the covariance histogram will become sufficiently smooth to allow us to estimate $P(\cov|\data)$ very well. The Blackwell-Rao estimator is the continuum limit of this idea; since we know the functional form of $P(\cov|\field)$ analytically, we can replace the covariance sampling step by the full analytical distribution. The density $P(\cov_\ell|\{\field_\ell\})$ is then estimated by,
\begin{align}
P(\cov_\ell|\{\field_\ell\})\propto\sum_i \f{|\bs{\Gamma}^i_\ell|^{(2\ell + 1)/2}}{|\cov_\ell|^{(2\ell+2+ p)/2}}\exp\left[-\f{1}{2}\mathrm{tr}(\cov_\ell^{-1}\boldsymbol{\Gamma}^i_\ell)\right].
\end{align}
This smooth density can then be used effectively for sampling $\bs{\theta}\sim P(\cov(\bs{\theta}, M)|\data)P(\bs\theta)/P(\cov(\bs\theta, M))$. In this work we draw samples from the posterior distribution of the tomographic shear maps and power spectra as a proof-of-concept for hierarchical map-power spectrum inference for weak lensing, but note that application of the Blackwell-Rao estimator for extracting cosmological inferences (given samples from the map-power spectrum posterior) is technically straightforward. For discussion of the Blackwell-Rao estimator in the context of CMB power spectrum inference see e.g., \citet{Wandelt2004}, and for application to CMB data see e.g., \citet{Chu2005}.

\subsection{Intrinsic Alignments}
\label{sec:ia}
The intrinsic alignment (IA) of nearby galaxies, arising from their evolution in a shared tidal gravitational environment, is known to be an important systematic effect when extracting cosmological inferences from cosmic shear, i.e., from correlations of observed galaxy ellipticities across the sky \citep[see e.g.,][for detailed reviews]{Joachimi2015IA, Kirk2015IA, Kiessling2015IA}. A careful analysis might include a parameterized model for intrinsic alignments, jointly inferring cosmological and intrinsic alignment model parameters simultaneously and marginalizing over the latter. It is straightforward to incorporate a model for intrinsic alignments into the framework presented in this paper. In the presence of intrinsic alignments, the angular power spectrum inferred from observed galaxy shapes will have contributions from both gravitational lensing and intrinsic alignments, i.e., $\cov = \cov^\mathrm{GG}(\bs\theta, M) + \cov^\mathrm{IA}(\bs\theta, \bs\alpha_\mathrm{IA}, M)$, where we now distinguish between the cosmic shear power spectra $\cov^\mathrm{GG}(\bs\theta, M)$ (described in preceding sections) and the intrinsic alignment power spectra $\cov^\mathrm{IA}(\bs\theta, \bs\alpha_\mathrm{IA}, M)$ (containing all IA terms), which are functions of both the cosmology and an additional set of parameters $\bs\alpha_\mathrm{IA}$ characterising the intrinsic alignment model. The inference process described in \S \ref{sec:vanilla}-\ref{sec:messenger} achieves inference on the power spectrum $\cov$ without reference to any cosmological or intrinsic alignment model; the obtained posterior density $P(\cov|\data)$ describes the inference on the full angular power spectrum, containing (in principle) both lensing and IA contributions. The joint posterior for the cosmological and intrinsic alignment model parameters can then be straightforwardly sampled, via $P(\cov|\data)$, by drawing samples from
\begin{align}
\{\bs\theta, \bs\alpha_\mathrm{IA}\}\sim P(\cov(\bs\theta, \bs\alpha_\mathrm{IA}, M)|\data)\frac{P(\bs\theta, \bs\alpha_\mathrm{IA})}{P(\cov(\bs\theta, \bs\alpha_\mathrm{IA}, M))},
\end{align}
analogously to Eq. \eqref{cosmo_posterior} and the description in \S \ref{sec:br}. Having sampled the joint posterior $P(\bs\theta, \bs\alpha_\mathrm{IA}|\data)$, the intrinsic alignment nuisance parameters $\bs\alpha_\mathrm{IA}$ can be easily marginalized over as desired. Performing power spectrum inference independent of any cosmological or intrinsic alignment model allows us to do parameter inference for different cosmological and IA models quickly and easily \emph{a posteriori}, directly from the posterior distribution of $\cov$, without the need to re-analyse the full dataset. This is a major advantage for the method presented here.

\begin{figure*}
\includegraphics[width = 17cm]{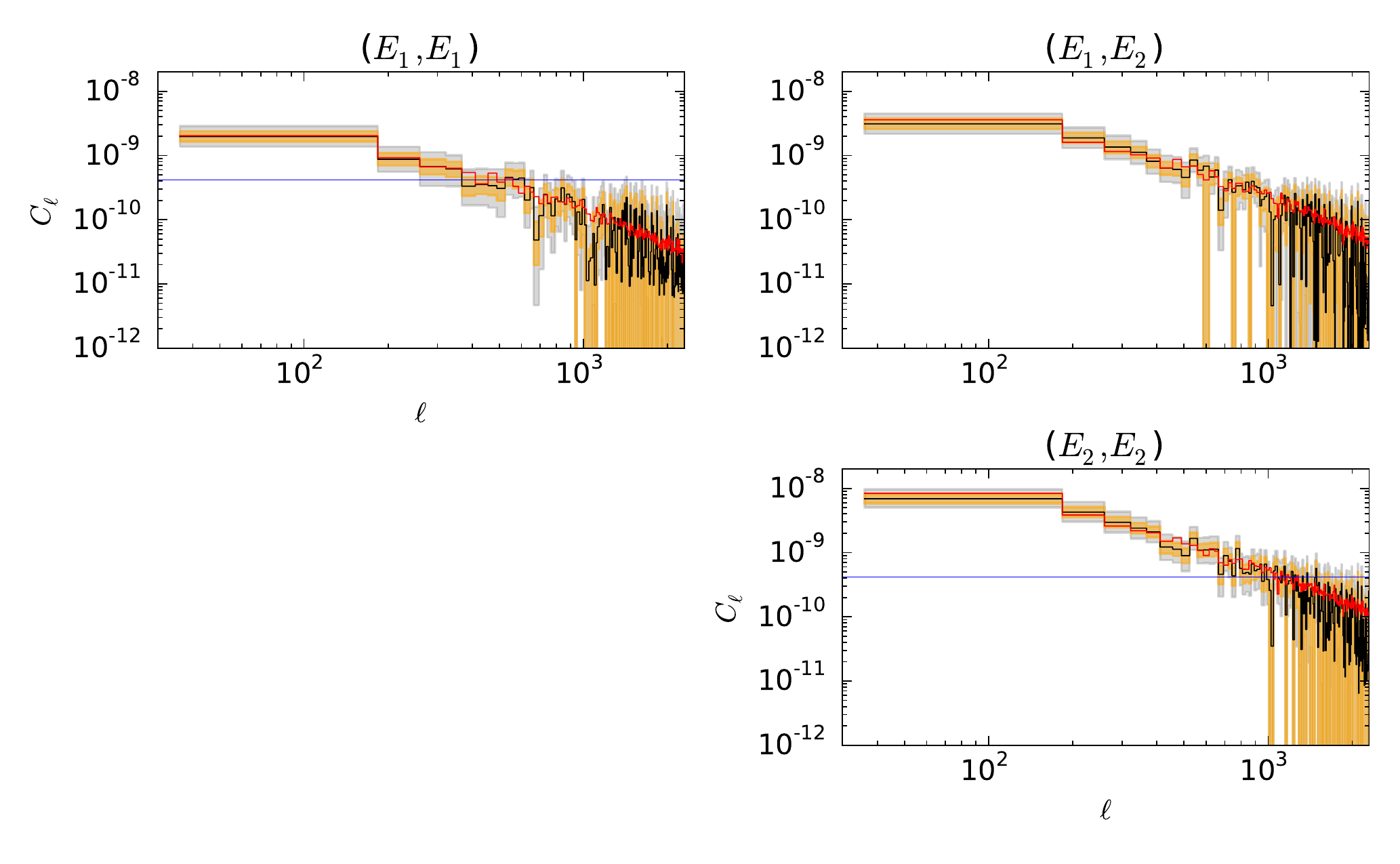}
\caption{Recovered $E$-mode tomographic shear power spectra: the inner orange bands indicate the $68\%$ credible region, the outer grey bands indicate the $95\%$ credible region, the black lines show the mean of the posterior distributions of the band powers, the red lines show the estimated band powers from the noiseless, mask-less simulated shear catalogue and the horizontal blue lines show indicate the mean ellipticity-noise level.}
\label{fig:ee}
\end{figure*}

\begin{figure*}
\includegraphics[width = 17cm]{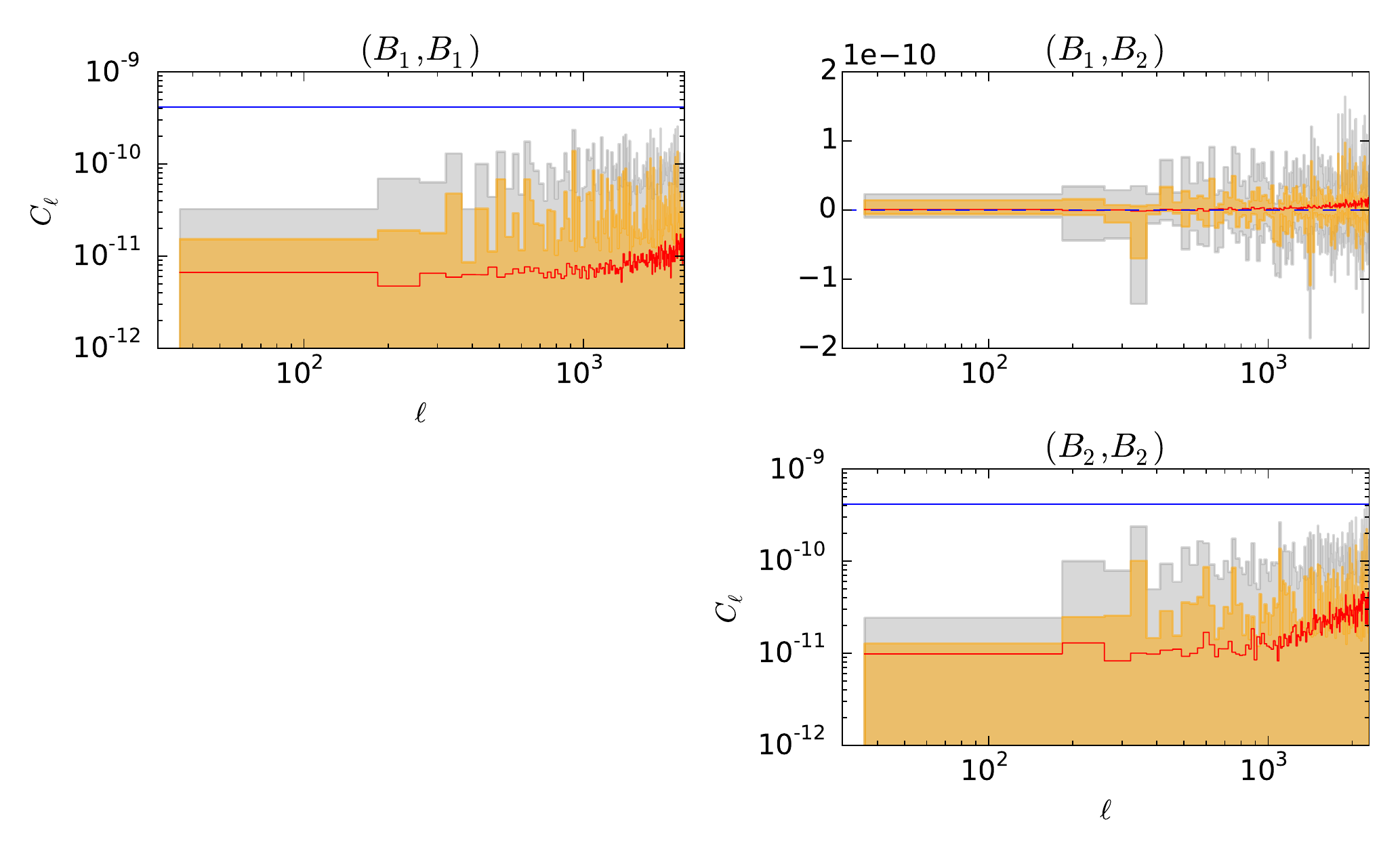}
\caption{Recovered $B$-mode tomographic shear power spectra: the inner orange bands indicate the $68\%$ credible region, the outer grey bands indicate the $95\%$ credible region, the red lines show the estimated band powers from the noiseless, mask-less simulated shear catalogue and the horizontal blue lines show indicate the average ellipticity-noise level. The recovered $B$ mode power spectra are consistent with the small $B$-mode signal in the simulated data. For the small amplitude of $B$-modes in the simulation ($1$--$2$ orders of magnitude below the ellipticity noise level), the posteriors are also consistent with zero, effectively providing upper limits only. In order to make stronger statements about the $B$-mode power at this low signal-to-noise, one must either motivate a more informative prior or a more informative model for the $B$-modes.}
\label{fig:bb}
\end{figure*}

\begin{figure*}
\includegraphics[width = 17cm]{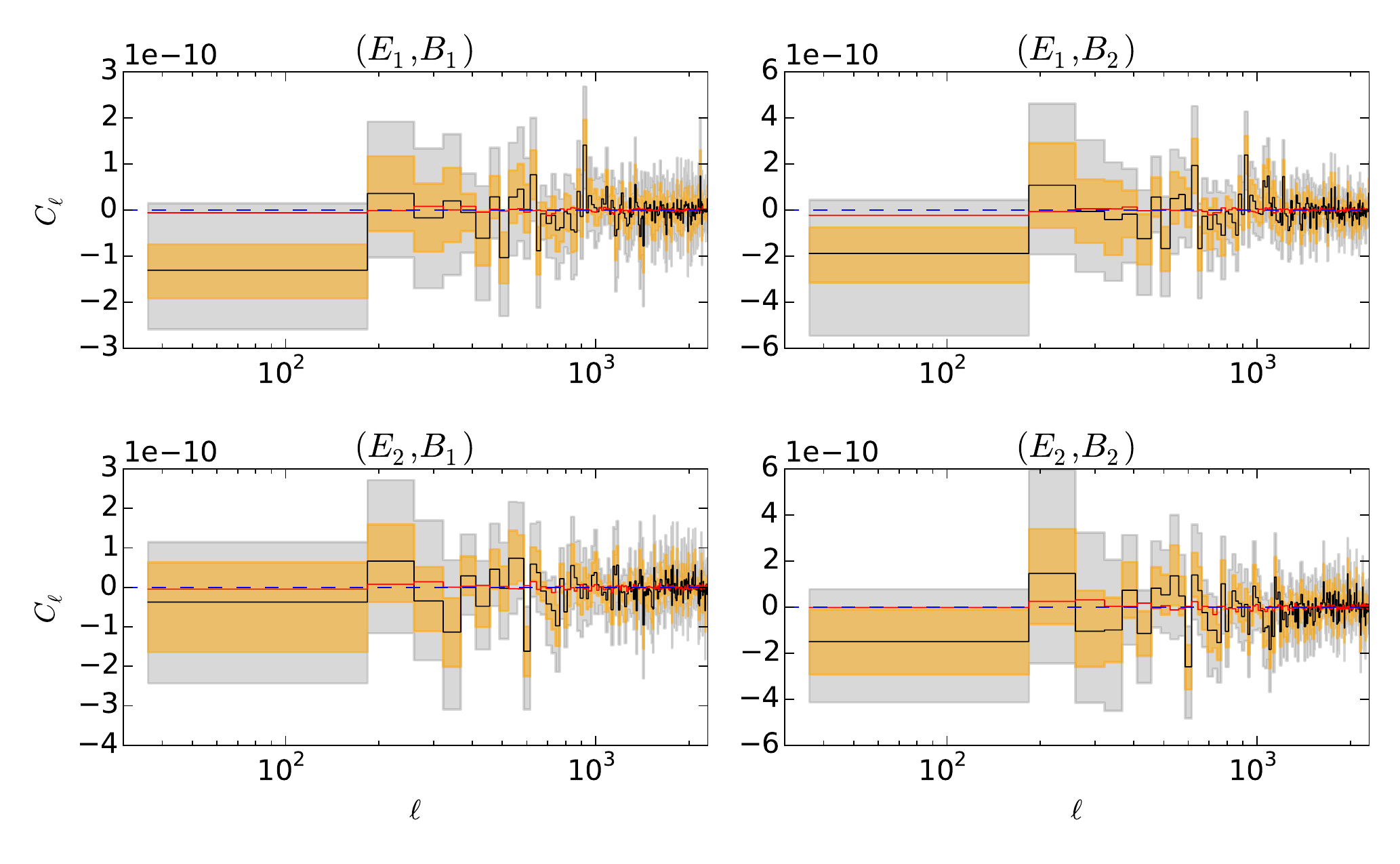}
\caption{Recovered $EB$-cross tomographic shear power spectra: the inner orange bands indicate the $68\%$ credible region, the outer grey bands indicate the $95\%$ credible region, the black lines show the mean of the posterior distribution of the band powers and the red lines show the estimated band powers from the noiseless, mask-less simulated shear catalogue. All recovered $EB$ cross powers are consistent with zero, as expected.}
\label{fig:eb}
\end{figure*}

\begin{figure*}
\includegraphics[width = 17cm]{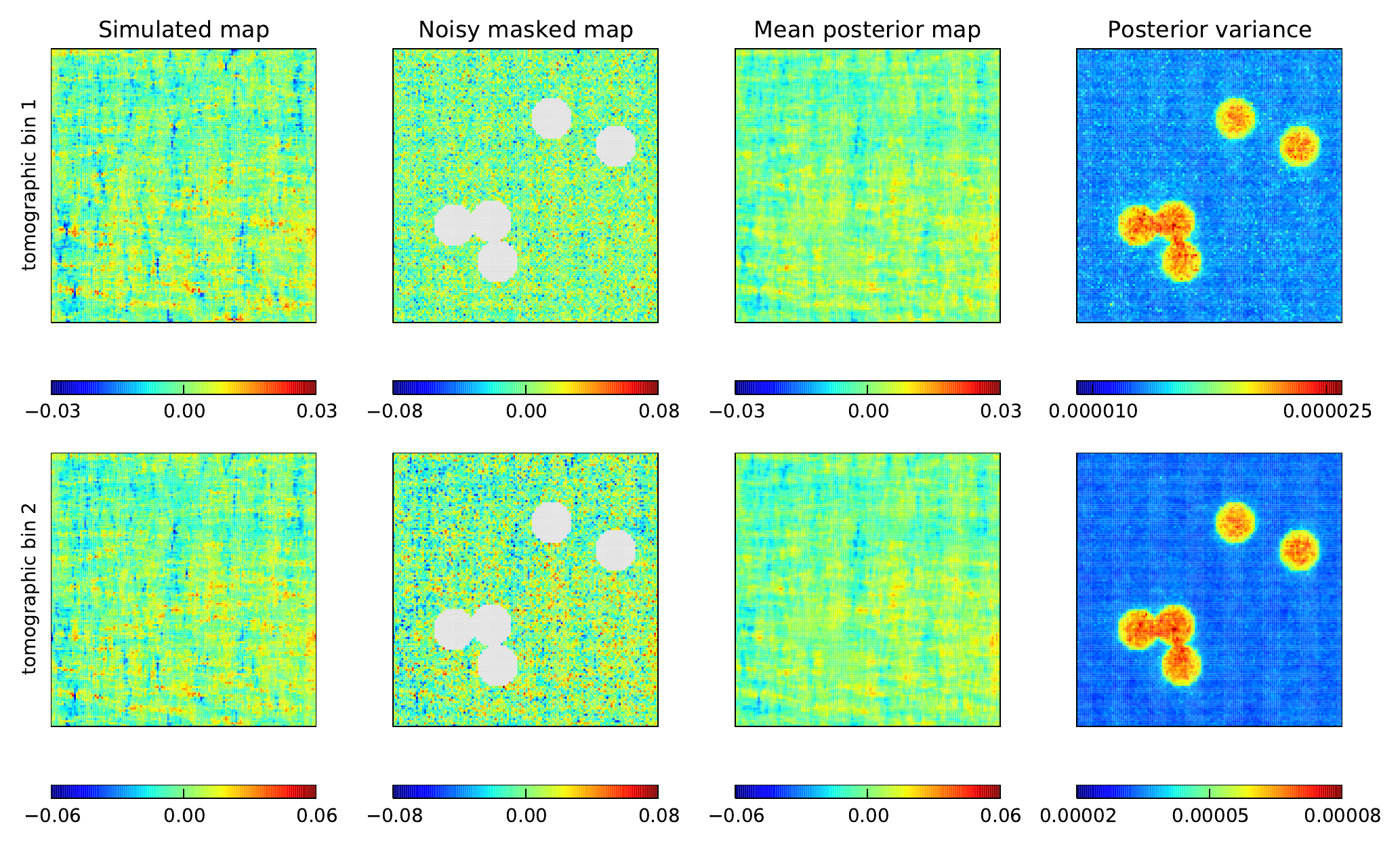}
\caption{Tomographic maps of the $\gamma_1$ component of the shear for the noiseless simulated shear maps (far left), noisy masked simulated maps (second from left), mean posterior maps (third from left) and the posterior variance (far right). The mean posterior maps recover most of the structure from the simulated shear maps. Note that inference is made about the field in masked regions, but the posterior variance in those regions is significantly higher than in unmasked regions as expected.}
\label{fig:maps}
\end{figure*}

\section{Simulations}
\label{sec:sims}
To demonstrate the joint map-power spectrum inference described in \S \ref{sec:map_power_inference} (including the messenger field sophistication), we apply the technique to simulated data. We generated a simulated shear catalogue using the \textsc{sunglass} weak lensing simulation pipeline \citealp{Kiessling2011}. \textsc{sunglass} generates cosmological $N$-body simulations with $512^3$ particles and a box of size length $512h^{-1}\mathrm{Mpc}$ (with a mass resolution of around $7.5\cdot 10^{10}M_\odot$) given a fixed $\Lambda$CDM cosmology using the \textsc{gadget}$2$ $N$-body code \citep{Springel2005}. It then computes weak lensing effects along a lightcone by performing line-of-sight integrations in the Born approximation (with no radial binning). The computed weak lensing fields are then interpolated back onto the particles in the lightcone, generating a mock 3D shear catalogue (see \citealp{Kiessling2011} for more details). We take a WMAP$7$ $\Lambda$CDM cosmology \citep{Larson2011, Jarosik2011} with $\Omega_\Lambda = 0.73$, $\Omega_m = 0.27$, $\Omega_b = 0.045$, $n_s = 0.96$, $\sigma_8 = 0.8$ and $h = 0.71$. The galaxy redshift distribution is given by $p(z) \propto z^2 e^{-(z/z_0)^{1.5}}$, with $z_0 = 0.7$ corresponding to a median redshift of around $1$, the lightcone extends to a redshift of $2$ and covers an area of $10\mathrm{deg}\times10\mathrm{deg}$. Gaussian photometric redshift errors are added to each galaxy redshift, with zero mean and dispersion $\sigma_z = 0.05(1+z)$. Each galaxy is also given an intrinsic random ellipticity, where the two ellipticity components are independently drawn from a Gaussian with zero mean and dispersion $\sigma_\epsilon = 0.27$. Note that realistic ellipticity distributions are not expected to be Gaussian. However, since many galaxies are averaged together to give the estimated shear in each pixel, our results will be insensitive to the shape of the assumed intrinsic ellipticity distribution, depending only on the variance of the random ellipticity components due to the central limit theorem (in this study, $\sim 330$ galaxies contribute to the estimated shear in each pixel and tomographic bin). The result is a 3D catalogue of galaxies with angular positions, photometric redshifts and observed ellipticities (given by the sum of the random intrinsic ellipticity and the shear for each source), covering $10\mathrm{deg}\times10\mathrm{deg}$, a redshift range from $z = 0$ to $z=2$, with an average of $30$ galaxies per square arc minute on the sky.

In order to perform a tomographic shear analysis, the catalogue is divided into two photometric redshift bins, with $0 < z_\mathrm{ph} \leq 1$ and $1 < z_\mathrm{ph} \leq 2$ respectively. The sources in each redshift bin are then grouped into $128\times 128$ angular pixels and the estimated shear in each pixel is computed as the mean of the galaxy ellipticities in each of the pixels. A survey mask is added by masking out $5$ circular patches positioned randomly over the survey area, each with radius $\sim 0.8\mathrm{deg}$. The result is two 2D tomographic $10\mathrm{deg}\times10\mathrm{deg}$ noisy shear maps with a non-trivial survey mask, as described in \S \ref{sec:data_vector}.
\section{Results}
\label{sec:results}
We applied the algorithm described in \S \ref{sec:messenger} and Algorithm \ref{algorithm:algo} to the simulated noisy tomographic shear maps generated using the \textsc{sunglass} simulation pipeline described in \S \ref{sec:sims}. Since the simulated patch of sky is small we employ the flat-sky approximations described in \S \ref{sec:flat}. It is highly desirable for the Gibbs sampler to bin several power spectrum coefficients together \citep{Larson2007, Eriksen2006}. This improves the sampling efficiency in two ways: firstly, the number of model parameters is reduced bringing down the computational cost of each Gibbs sampling step. Secondly, in the implementation described in \S \ref{sec:map_power_inference} the typical step size between two consecutive Gibbs samples is determined by the cosmic variance, whereas the full posterior has both cosmic variance and noise. As such, in the low signal-to-noise regime the sampler must make a large number of steps to obtain two independent samples of the power spectrum coefficients. By binning $\ell$ modes together into a set of band-power coefficients, we are effectively increasing the signal-to-noise of the now reduced set of power spectrum coefficients, improving the sampling efficiency. To this end, $\ell$ modes were binned together so that each bin contains $\geq 80$ modes, giving a total of $194$ band-power coefficients. Since there are two tomographic bins and two degrees-of-freedom for shear, this results in a total of $\num[group-separator={,}]{1940}$ model parameters for the power spectrum coefficients. The two tomographic shear maps contain $128\times 128$ pixels, each with two degrees-of-freedom, so the pixels constitute a further $\num[group-separator={,}]{65536}$ model parameters, bringing the total number of parameters in the inference task to $\num[group-separator={,}]{67476}$.

\begin{figure}
\begin{center}
\includegraphics[width = 9cm]{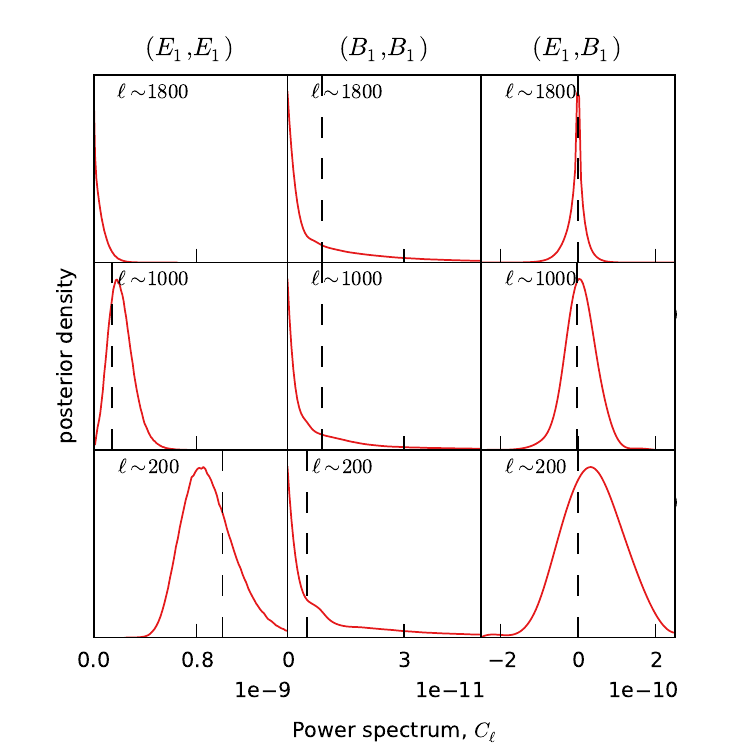}
\end{center}
\caption{Selected obtained posterior probability distributions for power spectrum coefficients for the $(E_1,E_1)$ (left column), $(B_1, B_1)$ (middle column) and $(E_1, B_1)$ (right column) powers at $\ell \sim 1800$ (top row), $\ell \sim 1000$ (middle row) and $\ell \sim 200$ (bottom row). The displayed posterior distributions are obtained from the smoothed histograms of posterior samples. The $E$-mode posterior densities are clearly negatively skewed, becoming increasingly asymmetric with increasing $\ell$ as the signal-to-noise per mode decreases.}
\label{fig:posteriors}
\end{figure}

We ran three Gibbs chains with independent starting points, each with $1.2$ million steps. Convergence was determined using the Gelman-Rubin statistic $r$ \citep{Gelman1992}; the chains were deemed to have converged to the marginal distributions for all parameters, with $r < 1.1$ in all cases. Each sampling step required a clock-time of $\sim0.5$ seconds on a high-end $2015$ desktop CPU. The obtained samples from the posterior distribution for the tomographic shear maps and power spectra are summarised in Fig. \ref{fig:ee}-\ref{fig:eb}; the inner orange bands show the $68\%$ credible regions, the outer grey bands show the $95\%$ credible regions, the black lines show the mean posterior band powers, the red lines show the estimated band powers from the noiseless (mask-less) simulated shear catalogue and the horizontal blue lines show the (average) pixel noise level due to random galaxy ellipticities. The mean and variance of the posterior distribution of the shear maps are shown alongside the simulated maps in Fig. \ref{fig:maps}. Obtained (smoothed) posterior distributions for the power spectrum coefficients for selected $\ell$-modes are shown in Fig. \ref{fig:posteriors}.

Fig. \ref{fig:ee} shows the recovered $E$-mode tomographic (cross) power spectra for the two tomographic bins; the posterior distributions for the power spectra summarized by the orange (68\%) and grey (95\%) credible regions and black line (posterior mean) are clearly recovering the underlying power spectra in the simulation (red line). The posterior mean follows the underlying simulated power (albeit with some scatter). Since the number of modes in each $\ell$-bin are chosen to be roughly the same, the cosmic variance is roughly the same for all of the band power coefficients. The increase in the posterior width with increasing $\ell$ is hence due entirely to the decreasing signal-to-noise, since the amplitude of the $E$-mode power spectra decreases as a function of $\ell$ whilst the noise remains constant (i.e., independent of scale).

Fig. \ref{fig:bb} shows the recovered $B$-mode tomographic power spectra. The posterior distributions for the $B$-mode power (for the auto-bins) are consistent with the small signal in the simulated data. Due to the small amplitude of the $B$-mode power (a factor of $10$--$100$ below the ellipticity noise level) the $B$-mode posteriors are also consistent with zero, effectively providing upper limits only. In order to make stronger statements about the $B$-modes (at this low signal-to-noise), one must either motivate a more informative prior or choose a more informative model for the $B$-mode power. In the absence of prior information or understanding of the $B$-mode signal, inference with an (uninformative) Jeffrey's prior, as chosen here, is appropriate. When zero $B$-modes are expected from cosmology and the $B$-mode power is used purely as a test of systematics, one could perform Bayesian model selection on a model with both $E$- and $B$-modes versus a model with $E$-modes only, in order to make stronger statements about the presence of residual $B$-modes in the data. The Bayesian hierarchical modelling approach provides a natural framework for such model selection. The recovered posterior inference on the $B$-mode cross power between the two tomographic bins is also consistent with the data (and consistent with zero).

Fig. \ref{fig:eb} shows the recovered $EB$ cross tomographic power spectra for the two tomographic bins. We expect zero correlation between $E$- and $B$-modes due to parity, and indeed all of the recovered $EB$ power spectra are consistent with zero as expected.

Fig. \ref{fig:maps} shows the mean and variance of the posterior distributions for the maps for the $\gamma_1$ component of the complex shear, alongside the noiseless (mask-less) simulated shear maps and the simulated maps with noise and mask added. The mean posterior maps are clearly recovering the structure in the underlying simulated shear maps. Importantly, note that inference of the shear field in the masked regions is obtained as described in \S \ref{sec:map_power_inference}, but the posterior variance in these regions is significantly higher than in the unmasked regions as one would expect, since the data provides no direct information about the field in those pixels; information is only obtained through the power spectrum inference (which provides prior information on the field at each sampling step) combined with the inferred field in pixels surrounding the masked regions.

Fig. \ref{fig:posteriors} shows the obtained posterior distributions for the $C_\ell$ coefficients for selected $\ell$-modes for $E$-mode, $B$-mode and $EB$-cross power (for the first tomographic bin only). For the $E$-modes, the posterior distributions are clearly non-Gaussian showing significant negative skewness; the method has captured the full non-Gaussian shape of the posterior distributions of the power spectrum coefficients. The asymmetry increases with increasing $\ell$ as the signal-to-noise (per mode) decreases. For the $B$-modes, where the signal-to-noise is very low across the full range of $\ell$, the posteriors are peaked at (or close to) zero with positive tails; these will be somewhat prior dependent, although we note that at very low signal-to-noise this behaviour may be expected for a range of (weakly informative) priors. We reiterate that in order to make stronger statements about the $B$-modes at this low signal-to-noise, one must provide a (motivated) more informative prior or more informative model.

\section{Conclusions}
\label{sec:conclusions}
We have developed a Bayesian hierarchical approach to cosmic shear power spectrum inference, whereby we sample from the joint posterior distribution of the shear map and power spectrum in a Gibbs sampling framework. Similar methods have been developed and applied for power spectrum inference in the context of CMB temperature, polarization and large-scale structure problems. The joint map-power spectrum inference approach has a number of highly desirable features. Complicated survey geometry and masks are trivially accounted for in contrast to approximate power spectrum estimation methods such as the pseudo-$C_\ell$ approach, since the inference of the power spectrum and unmasked pixels provides prior constraints on the field in the masked regions where the data provides no information directly. The posterior distribution on the shear power spectrum is a powerful intermediate product for a cosmic shear analysis, since it allows cosmological parameter inference and model selection to be performed without loss of information for a wide range of models \emph{a posteriori}; this is clearly preferable to analysing the full data set for each model independently. The method is exact and optimal under the assumption of Gaussian fields, in the sense that no information is lost from the data once it has been pre-processed into estimated noisy shear maps. Loss of information and biases in this pre-processing step (from raw pixel data to estimated galaxy shapes and redshifts) can be avoided by embedding the hierarchical map-power spectrum model into a larger global hierarchical model for cosmic shear accounting for instrumental effects, modelling distributions of intrinsic galaxy properties and cosmology together in a single global analysis.

We demonstrate the method by performing tomographic map-power spectrum inference on a simulated shear catalogue of galaxies with random intrinsic ellipticities, distributed across the sky and in redshift (with photo-$z$ errors), processed into two tomographic, pixelized, noisy shear maps, with a complicated survey mask. The simulation covers $10\;\mathrm{deg}\times10\;\mathrm{deg}$, extends to redshift $z=2$ and contains on average $30$ galaxies per square arc minute. The map-power spectrum inference approach produces posterior distributions for the shear map and tomographic power spectra, successfully recovering the simulated $E$-mode, $B$-mode and $EB$-cross tomographic power spectra from the simulation.

The Bayesian map-power spectrum inference approach can be straightforwardly extended to 3D shear power spectrum inference, joint shear-magnification analysis \citep{Heavens2013, Alsing2015a} or joint analysis of weak lensing with other probes (galaxy clustering, CMB etc). Non-Gaussianity of the shear field on small angular scales means the Gaussian distribution for the shear field taken in this work will be sub-optimal when extracting inferences on these scales. The framework developed here can in principle be extended to jointly estimate the power spectrum and higher-order statistics for the shear field, capturing more of the available information on non-linear scales. This would require specifying a model for the non-Gaussian shear probability density; the appearance of this non-Gaussian probability density in the hierarchical model will likely make the conditional distributions too complicated to permit Gibbs sampling. In this case an alternative sampling scheme, such as Hamiltonian Monte Carlo (HMC), should be developed (see e.g., \citealp{Jasche2010a} for a successful application of non-Gaussian field inference using HMC in the context of large-scale structure). We will investigate these issues in future work.
\section*{acknowledgments}
We would like to thank Thomas Kitching, Guilhem Lavaux, Daniel Mortlock and others for stimulating discussions. AK was supported in part by JPL, run under contract by Caltech for NASA. AK was also supported in part by NASA ROSES 13-ATP13-0019. BW acknowledges funding from an ANR Chaire d'Excellence (ANR-10-CEXC-004-01), the UPMC Chaire Internationale in Theoretical Cosmology and the Labex Institut Lagrange de Paris (ANR-10-LABX-63) part of the Idex SUPER.

\bibliographystyle{mn2e_williams}
\bibliography{power_inference}

\end{document}